\documentclass[12pt,a4paper]{article}

\usepackage{amsfonts}
\usepackage{amsmath}
\usepackage{amssymb,bm}
\usepackage{latexsym}
\usepackage{amsthm, mathrsfs}
\usepackage{color}
\usepackage{subfig}
\usepackage[section]{placeins} 

\usepackage[dvips]{graphicx}

\numberwithin{equation}{section}

\usepackage{tikz}
\usetikzlibrary{decorations.pathmorphing}
\usetikzlibrary{decorations.markings} 
\usepgflibrary{arrows}

\newcommand{\BbbZ}{\mathbb{Z}}

\newcommand{\x}{\mathsf{x}}
\newcommand{\U}{\mathsf{U}}

\DeclareMathOperator{\expo}{e}
\newcommand{\scri}{\mathscr{I}}
\newcommand{\scrh}{\mathscr{H}}
\newcommand{\be}{\begin{equation}}
\newcommand{\ee}{\end{equation}}
\newcommand{\ba}{\begin{eqnarray}}
\newcommand{\ea}{\end{eqnarray}}
\newcommand{\bea}{\begin{equation}\begin{aligned}}
\newcommand{\eea}{\end{aligned}\end{equation}}
\newcommand{\baa}{\begin{array}}
\newcommand{\eaa}{\end{array}}

\newcommand{\bfx}{{\bf x}}

\DeclareMathOperator{\arcsinh}{arcsinh}

\DeclareMathOperator{\phin}{\phi^{\text{in}}_{\omega\ell}}
\DeclareMathOperator{\phup}{\phi^{\text{up}}_{\omega\ell}}
\DeclareMathOperator{\Phin}{\Phi^{\text{in}}_{\omega\ell}}
\DeclareMathOperator{\Phup}{\Phi^{\text{up}}_{\omega\ell}}

\DeclareMathOperator{\tPhin}{\tilde{\Phi}^{\text{in}}_{\omega\ell}}
\DeclareMathOperator{\tPhup}{\tilde{\Phi}^{\text{up}}_{\omega\ell}}

\DeclareMathOperator{\rin}{\mathsf{\rho^{\text{in}}_{\omega\ell}}}
\DeclareMathOperator{\rup}{\rho^{\text{up}}_{\omega\ell}}
\newcommand{\rinf}{r_{\text{inf}}}
\newcommand{\ninf}{n_{\text{inf}}}
\newcommand{\rh}{r_{H}}
\newcommand{\nh}{n_{H}}
\newcommand{\Aup}{A^{\text{up}}_{\omega\ell}}
\newcommand{\Ain}{A^{\text{in}}_{\omega\ell}}
\newcommand{\Bup}{B^{\text{up}}_{\omega\ell}}
\newcommand{\Bin}{B^{\text{in}}_{\omega\ell}}

\title{Static detectors and circular-geodesic detectors on the Schwarzschild black hole}

\author{Lee Hodgkinson${}^{1}$\thanks{lee.hodgkinson@nottingham.ac.uk}
\ ,
Jorma Louko${}^{1}$\thanks{jorma.louko@nottingham.ac.uk}
\ and
Adrian C. Ottewill${}^{2}$\thanks{adrian.ottewill@ucd.ie}
\\
\noalign{\vspace{3ex}}
\small{\it ${}^{1}$ School of Mathematical Sciences,
}\\
\small{\it University of Nottingham, Nottingham NG7 2RD, UK}
\\[1ex]
\small{\it ${}^{2}$ School of Mathematical Sciences and Complex \& Adaptive Systems Laboratory,}
\\
\small{\it University College Dublin, UCD, Belfield, Dublin 4, Ireland}
\\[2ex]
\small{(Revised March 2014)}
\\[1ex]
\small{Published in Phys.\ Rev.\ D {\bf 89}, 104002 (2014)}}

\date{}

\begin{document}

\maketitle

\begin{abstract}

We examine the response of an Unruh-DeWitt particle detector coupled to
a massless scalar field on the (3+1)-dimensional Schwarzschild
spacetime, in the Boulware, Hartle-Hawking and Unruh states, for static
detectors and detectors on circular geodesics, by primarily numerical
methods. For the static detector, the response in the Hartle-Hawking
state exhibits the known thermality at the local Hawking temperature,
and the response in the Unruh state is thermal at the local
Hawking temperature in the limit of a large detector energy gap. 
For the circular-geodesic
detector, we find evidence of thermality in the limit
of a large energy gap for the Hartle-Hawking and Unruh
states, at a temperature that exceeds the Doppler-shifted local Hawing
temperature. Detailed quantitative comparisons between the three
states are given. The response in the Hartle-Hawking state is
compared with the response in the Minkowski vacuum and in the
Minkowski thermal state for the corresponding
Rindler, drifted Rindler, and circularly accelerated trajectories. 
The analysis takes place within first-order perturbation theory 
and relies in an essential way on stationarity.

\end{abstract}

\newpage

\section{Introduction}
\label{sec:intro}

The Unruh-DeWitt detector~\cite{unruh, deWitt} is a simple model
particle detector --- a two-state quantum-mechanical system coupled to
the quantum field.  As we shall see, the particular quantum-mechanical
system chosen is unimportant; however, the reader may find it helpful
to keep in mind a concrete picture of the detector as, say, a
two-state hydrogen atom.

On a curved spacetime or for non-inertial observers, the particle
content of the field is ambiguous, and a distinguished notion of a
``particle'', defined with respect to some timelike Killing vector,
may not exist. Unruh-DeWitt detectors liberate us from the reliance on
symmetries to investigate particle content, and they provide an
operational definition of the field's particle content. The
tool of Unruh-DeWitt detectors has been applied in several situations,
which include accelerated observers in Minkowski
spacetimes \cite{unruh,deWitt,Letaw:1980yv,Grove:1983rp,takagi,Muller:1995vk,Crispino:2007eb},
static detectors in exterior
Schwarzschild \cite{Hartle:1976tp,Israel:1976ur}, inertial detectors
in de Sitter space~\cite{gibb-haw:dS}, and static, co-rotating and
freely-falling detectors on the Ba\~nados-Teitelboim-Zanelli (BTZ)
black hole~\cite{Hodgkinson:2012mr}.

To study an Unruh-DeWitt detector in a given spacetime, we use
first-order perturbation theory, and we are led to the notions of the
detector response function and the instantaneous transition
rate. Heuristically, the detector response function gives the
probability of a transition between the states of the detector, and
the transition rate represents the ``number of particles detected per
unit proper time''. The transition rate will be the primary quantity
of interest in this paper. In general, one must take extreme care in
obtaining the response function and transition rate
\cite{louko-satz:profile,satz:smooth,satz-louko:curved,Hodgkinson:2011pc};
sufficient conditions are that the quantum state of the field is
regular in the Hadamard sense \cite{kay-wald,Decanini:2005gt} and that
the detector is switched on (off) smoothly
\cite{Fewster:1999gj,junker,hormander-vol1,hormander-paper1}.

However,
in this paper we shall be concerned with stationary detectors for 
which the initial state of the field is invariant under the timelike 
Killing vector that generates the trajectory, and owing to this, 
the switch-on of the detector can be pushed to the infinite past and 
the transition rate is time-independent, within first-order perturbation 
theory. In this case, we can bypass the sensitive issues of 
regularisation and switching by 
integrating formally over the whole trajectory and factoring out 
the infinite total proper time integral~\cite{byd}.

In this paper, we use numerical methods to solve the radial 
part of the Klein-Gordon equation for a massless scalar field on 
four-dimensional Schwarzschild spacetime. We investigate static 
detectors and also detectors on circular geodesics 
(stable or unstable) in the exterior region, 
and we consider the Hartle-Hawking, Boulware and Unruh states for the field. 

For the static detector in the Hartle-Hawking state, the response is
known to be thermal \cite{Hartle:1976tp} in the Kubo-Martin-Schwinger
(KMS) sense \cite{Kubo:1957mj,Martin:1959jp} 
--- 
the local Hawking temperature at radial coordinate 
$R>2M$ is equal to $T_{\text{loc}}=T_H(1-2M/R)^{-1/2}$, 
where $M>0$ is the mass of the black hole and 
$T_H=(8\pi M)^{-1}$ is the asymptotic temperature at 
infinity~\cite{Tolman,hawking}. 
Our numerical calculation provides the detailed 
profile of this thermal response, including the factors due to the
density of states on the curved background~\cite{Ottewill:1987tm}.

For the circular-geodesic detectors in the Hartle-Hawking and Unruh
states, we find evidence that the detector responds thermally in the
limit of a large energy gap, at a temperature higher than that recorded
by the static detector, by a factor larger than the 
time dilation Doppler shift factor. 
We show analytically that the same phenomenon occurs 
for a stationary detector in three qualitatively similar
situations in Minkowski spacetime. 
The physical explanation for the blueshift exceeding time
dilation appears to be that 
the transition rate at large excitation energies is dominated 
by the most energetic field quanta, and these are seen by the detector 
from a head-on direction and are hence blueshifted 
more than just by time dilation. 
This explanation is consistent with the analysis 
of a circular-geodesic detector in \cite{Smerlak:2013sga}
within a model in which the 
angular dependence of the field is suppressed, where 
the asymptotic temperature in a state closely resembling the Unruh state 
was found to be related 
to the local Hawking temperature by just the time 
dilation Doppler shift factor. 

We see no qualitative difference between the transition 
rates of stable versus unstable circular orbits. 

For both the static and circular-geodesic detectors, we 
find that the Boulware and Unruh rates align at negative 
detector energy gap when the distance from the hole is large. 
This is consistent with the Unruh state mimicking an outgoing 
flux of radiation from a collapsing star that diminishes as 
$r^{-2}$ and the fact that the Boulware state reduces to the 
Minkowski vacuum at spatial infinity. Similarly, 
because of this property of the Unruh state and the fact the 
Hartle-Hawking state represents a thermal bath at infinity, 
we see that the ratio of the excitation part of 
the transition rate in the Hartle-Hawking state to the 
transition rate in the Unruh state becomes large as the radius increases.

Considering the static detector, as the magnitude of the energy gap 
is increased further, all three states align for large negative energies.
For both the static and 
circular-geodesic detectors, we see that at large magnitude 
energy gaps, the transition rate becomes oscillatory when the 
detector is near the horizon; this effect is due to 
back-scattering by the effective potential induced by the 
curvature. It is reminiscent of the oscillation observed 
for the BTZ black hole in~\cite{Hodgkinson:2012mr}, 
and we verify analytically that similar oscillations take place also 
in Minkowski space for a field with an external potential barrier. 

For the static detector with the field in the Boulware state, we see
that the transition rate has a vanishing excitation rate (only de-excitation), 
and this is consistent with the fact that for a static
trajectory we are on an orbit of the Schwarzschild time-translation
Killing vector. Whereas for the circular detector with the field in
the Boulware state, the transition rate now has a small excitation
component due to synchrotron radiation~\cite{Grove:1983rp}, and this
component gets smaller as the radius from the hole increases ---
consistent with the fact the circular detector is asymptotically
static.

When the field is in the Unruh state, we find that the transition 
rate of the static and circular-geodesic detectors at a given radius 
remains non-vanishing even in the limit of a vanishing 
excitation energy. This may at first seem surprising: 
one may expect that such low-frequency modes in the outgoing flux, 
which the Unruh state simulates, would not be able to escape the 
gravitational potential. We address these concerns by verifying 
the non-vanishing result analytically. 

We investigate the analogy between a uniformly linearly accelerated 
detector coupled to a field in the Minkowski vacuum in flat spacetime 
with the static detector in the Hartle-Hawking state on the 
Schwarzschild black hole, finding that the transition rates 
in the two cases align --- the agreement becoming better 
as the radius increases. Similarly, for the circular-geodesic 
detector a comparison between its transition rate and that of a 
Rindler detector given a transverse drift velocity is analysed, 
and we find good agreement as the radius increases. Making the drift 
direction periodic does not, however, improve the agreement.

The plan of the paper now follows. In Section \ref{sec:techIntro} we
give an overview of the Unruh-DeWitt detector model, 
introducing the usual notion of a transition rate for stationary situations within 
first-order perturbation theory. In
Section \ref{sec:4DSchw} we recall relevant features 
of the massless Klein-Gordon equation 
in four-dimensional Schwarzschild spacetime: 
in particular, we present in Section \ref{sec:4DSchw:bcs} our 
numerical scheme to solve the radial equation under 
the appropriate boundary conditions 
and in Section \ref{sec:4DSchw:norm} our procedure for 
normalising the numerical solutions. 
Section \ref{sec:WhichVac} presents the mode sum expressions for the 
Wightman function in the 
Hartle-Hawking state, the Boulware state and the Unruh state. 
Section \ref{sec:4DSchw:static}
presents the mode sum expressions for the transition rate 
for a static detector in each of these tree quantum states, 
and Section \ref{sec:4DSchw:circ} presents the similar 
mode sum expressions for a detector on a circular geodesic. 
Section \ref{sec:4DSchw:compRind} sets up the analytic framework for comparing the 
static and circular-geodesic detectors in Schwarzschild to 
detectors in Rindler space, respectively on worldlines of 
uniform linear acceleration and on worldlines of uniform linear acccleration 
with a transverse drift. 

The numerical results are presented and analysed in
Section~\ref{sec:4DSchw:results}. Section \ref{sec:4DSchw:discussion}
gives a summary and concluding remarks.  The three appendices analyse
respectively the small energy behaviour of the Schwarzschild radial
mode functions, a static detector coupled to a field with an external
potential barrier in Minkowski spacetime, and a detector in Minkowski
spacetime in stationary situations where there is a nonvanishing drift
velocity with respect to thermally responding detectors.

Our metric signature is $({-}{+}{+}{+})$. We use units in which
$c=\hbar= k_B = 1$, so that frequencies, energies and 
temperatures have dimension inverse length. 
The Schwarzschild mass parameter 
is denoted by $M$ and has dimension length. 
Spacetime points and Lorentz four-vectors are denoted with
sans-serif letters~($\x$), and Euclidean three-vectors are denoted
with bold letters~($\bfx$). For the Minkowski or Euclidean product of
two vectors of the respective kind we use a dot notation, $\x\cdot\x$
or $\bfx\cdot\bfx$.
%
%
\section{Unruh-DeWitt detectors in static or stationary settings}
\label{sec:techIntro}
In this section, we give a brief overview of the Unruh-DeWitt detector
model \cite{unruh,deWitt}. 
The Unruh-DeWitt detector is an idealised `atom'; 
it is spatially point-like and comprises two states: 
$|0_d\rangle$, which has energy $0$, and the state $|E_d\rangle$, 
which has energy~$E$, where $E$ may be a positive or negative real number.
\par
The detector moves through spacetime on the trajectory $\x(\tau)$,
where $\tau$ is the detector's proper time.  This simple
quantum-mechanical system (detector) is coupled to a real, free,
scalar, quantum field, $\phi$, by the interaction Hamiltonian
\be
H_{\text{int}}=c\chi(\tau)\mu(\tau)\phi(\x(\tau))\,,
\label{eq:techIntro:Hint} 
\ee
where $c$ is a small coupling-constant, $\chi$ is known as the
switching-function and $\mu$ is the monopole-moment operator of our
`atom'. We can think of the switching function $\chi$ as turning on
(off) our detector; in other words, as $\chi$ goes to zero the
detector and field are decoupled, so no particles in the field are
detected. In a general situation, we would now emphasise the necessity
of $\chi$ being smooth and of compact support, but in the case of
trajectories that are generated by the timelike Killing vectors of the
spacetime, this requirement can be relaxed.
\par 
During the course of the detector's motion through spacetime, the
detector will absorb (emit) quanta of energy, (de-)exciting it from
its initial state to alternative state. The first question we must
address is ``what is the probability of such a transition occurring?''
We answer this question within the framework of first-order
perturbation theory.
\par 
We shall assume the field is initially in some arbitrary Hadamard state~\cite{kay-wald}. Hadamard states have many desirable properties. In a Hadamard state, the stress-energy tensor is guaranteed to be renormalisable, and the singularity structure of the Wightman function in the coincidence limit is well defined~\cite{kay-wald}. All the states considered in this paper (Hartle-Hawking, Boulware and Unruh states on the Schwarzschild black hole) are Hadamard states.
\par 
We shall denote this initial Hadamard state of the field as $|\Psi\rangle$, and before the interaction begins, we assume the detector to be in the state $|0\rangle_d$. The detector-field system is, hence, initially in the composite state $|0\rangle_d\otimes|\Psi\rangle$. Regardless of the final state of the field, we are interested in the probability for the detector to be found in the state $|E\rangle_d$ after the interaction has ceased. Working in first-order perturbation theory, this probability factorises as~\cite{byd,wald-smallbook}
\begin{equation}
P(E)=c^2{|_d\langle0|\mu(0)|E\rangle_d|}^2\mathcal{F}\left(E\right)
\ , 
\label{eq:techIntro:prob}
\end{equation}
where the response function $\mathcal{F}(E)$ 
encodes the information about the detector's trajectory, 
the initial state of the field and the way the detector has been switched on and off.
$\mathcal{F}(E)$~can be expressed as
\begin{equation}
\mathcal{F}(E)=
\lim_{\epsilon\to0_+} 
 \int_{-\infty}^{\infty}\mathrm{d}\tau^{\prime} \,
\int_{-\infty}^{\infty}\mathrm{d}\tau^{\prime\prime} \,\chi(\tau^{\prime})\chi(\tau^{\prime\prime})\, 
\mathrm{e}^{-iE (\tau^{\prime}-\tau^{\prime\prime})}\, W_\epsilon(\tau^{\prime},\tau^{\prime\prime})\,,
\label{eq:techIntro:respfunc}
\end{equation}
where $W_\epsilon(\tau^{\prime},\tau^{\prime\prime})$ is a
one-parameter family of functions that converge to the pull-back of
the Wightman distribution on the detector's
worldline~\cite{satz-louko:curved,kay-wald,Fewster:1999gj,junker}. The
factors in front of $\mathcal{F}(E)$ in~\eqref{eq:techIntro:prob}
depend only on the internal structure of the detector and we shall
from now on drop them, referring to $\mathcal{F}(E)$ as the transition
probability.
\par
If we restrict the motion of the detector to be along the orbit of a
timelike Killing vector, and we assume $|\Psi\rangle$ to be invariant
under the isometry generated by this Killing vector, the Wightman
function is time-translation invariant along the trajectory, and we
are free to push the switch-on time of the detector to the asymptotic
past --- effectively replacing the switching-function $\chi$ by the
Heaviside step-function. With a change of variables, the result is
that we can formally just drop the external $\tau^{\prime}$-integral
of~\eqref{eq:techIntro:respfunc} in order to obtain the
\emph{transition rate}~\cite{byd}:
\begin{equation}
\mathcal{\dot{F}}\left(E\right)= \lim_{\epsilon\to0_+} 
\int^{\infty}_{-\infty}\,\mathrm{d}s\,\mathrm{e}^{-iEs}\, W_{\epsilon}(s)\,.
\label{eq:techIntro:transRate}
\end{equation}
The transition rate represents the number of particles detected per
unit proper time (this interpretation is slightly simplistic,
see~\cite{satz-louko:curved}), and it will be the primary quantity of
interest throughout this paper.
\section{Schwarzschild antecedents}
\label{sec:4DSchw}
In four-dimensional Schwarzschild spacetime, the Wightman function is
not known analytically, and in this section we present the numerical
methods necessary to study the transition rate of a detector coupled
to a massless, minimally-coupled, scalar field in the Hartle-Hawking,
Boulware and Unruh states.
\par
The metric of the Schwarzschild spacetime is given by
\begin{equation}
ds^2=-\left(1-\frac{2M}{r}\right)dt^2+\left(1-\frac{2M}{r}\right)^{-1}dr^2+r^2\left(d\theta^2+\sin^2{\theta}d\phi^2\right)\,,
\label{eq:4DSchw:metric}
\end{equation}
where we assume the mass parameter $M$ to be positive, the black hole
exterior is covered by $2M<r<\infty$, and the horizon is at $r\to 2M$.
\par
Mode solutions of the Klein-Gordon equation in the Schwarzschild spacetime have the form~\cite{byd}
\begin{equation}
\frac{1}{\sqrt{4\pi\omega}}r^{-1}\rho_{\omega\ell}(r)Y_{\ell m}\left(\theta,\phi\right)\expo^{-i\omega t}\,,
\label{eq:4DSchw:modeSolns}
\end{equation}
where $\omega>0$, $Y_{\ell m}$ is a spherical harmonic and the radial function $\rho_{\omega\ell}$ satisfies 
\begin{equation}
\frac{d^2 \rho_{\omega\ell}}{dr^{*2}}+\left\{\omega^2-\left(1-\frac{2M}{r}\right)\left[\frac{\ell(\ell+1)}{r^2}+\frac{2M}{r^3}\right]\right\}\rho_{\omega\ell}=0 \,,
\label{eq:4DSchw:radModTort}
\end{equation}
with $r^{*}$ being the tortoise co-ordinate, defined as
\begin{equation}
r^{*}=r+2M\log{\left(r/2M-1\right)} \,.
\label{eq:Schw:4d:tortoise}
\end{equation}
\par Alternatively, one can work with the Schwarzschild radial co-ordinate $r$ and define the function $\phi_{\omega\ell}(r):=\rho_{\omega\ell}(r)/r$, which satisfies
\begin{equation}
\phi^{\prime\prime}_{\omega\ell}(r)+\frac{2\left(r-M\right)}{r\left(r-2M\right)}\phi^{\prime}_{\omega\ell}(r)+\left(\frac{\omega^2 r^2}{(r-2M)^2}-\frac{\lambda}{r(r-2M)}\right)\phi_{\omega\ell}(r)=0\, ,
\label{eq:4DSchw:radModPhi}
\end{equation}
with $\lambda:=\ell(\ell+1)$. Solutions to
neither~\eqref{eq:4DSchw:radModTort} nor~\eqref{eq:4DSchw:radModPhi}
can be found analytically, and as such we seek the solutions
$\phi_{\omega\ell}(r)$ numerically using code written in Mathematica
(TM)~\cite{Mathematica}.

\begin{figure}[t] \centering
\tikzset{->-/.style={decoration={
  markings,
  mark=at position #1 with {\arrow[scale=1]{>}}},postaction={decorate}}}

\begin{center}
\begin{tikzpicture}[>=open triangle 90, scale=0.7]
\node (I)    at ( 4,0) {};
\node (II)   at (-5,0) {};

\path 
   (I) +(90:4)  coordinate[label=90:$i^+$]  (Itop)
       +(-90:4) coordinate[label=-90:$i^-$] (Ibot)
       +(180:4) coordinate (Ileft)
       +(0:4)   coordinate[label=0:$i^0$] (Iright)
       +(45: 2.825)   coordinate (Imidtr)
       +(225:2.825)   coordinate (Imidbl)
       +(135:2.825)   coordinate (Imidtl)
       +(-45:2.825)   coordinate (Imidbr)
       ;
       ;
\draw  (Ileft)  -- 
          node[midway, below, sloped] {$H^+$}
       (Itop)   -- 
          node[midway, above right]    {$\scri^+$}
          node[midway, below, sloped] {$\bar{v}=\infty$}
       (Iright) --
          node[midway, below right]    {$\scri^-$}
          node[midway, above, sloped] {$\bar{u}=-\infty$}
       (Ibot)   --
          node[midway, below, sloped]    {$H^-$}
       (Ileft)  -- cycle;
       
\draw[->-=.5] (Imidbr.center) -- (I.center);
\draw[->-=.5] (I.center) -- (Imidtl.center);   
\draw[->-=.5] (I.center) -- (Imidtr.center);    

\path 
   (II) +(90:4)  coordinate[label=90:$i^+$]  (IItop)
       +(-90:4) coordinate[label=-90:$i^-$] (IIbot)
       +(180:4) coordinate (IIleft)
       +(0:4)   coordinate[label=0:$i^0$] (IIright)
       +(45: 2.825)   coordinate (IImidtr)
       +(225:2.825)   coordinate (IImidbl)
       +(135:2.825)   coordinate (IImidtl)
       +(-45:2.825)   coordinate (IImidbr)
       ;
\draw  (IIleft)  -- 
          node[midway, below, sloped] {$H^+$}
       (IItop)   -- 
          node[midway, above right]    {$\scri^+$}
          node[midway, below, sloped] {$\bar{v}=\infty$}
       (IIright) --
          node[midway, below right]    {$\scri^-$}
          node[midway, above, sloped] {$\bar{u}=-\infty$}
       (IIbot)   --
          node[midway, below, sloped]    {$H^-$}
       (IIleft)  -- cycle;

\draw[->-=.5] (IImidbl) -- (II.center);
\draw[->-=.5] (II.center) -- (IImidtl);   
\draw[->-=.5] (II.center) -- (IImidtr);

\end{tikzpicture}
\caption{Illustrating the `up' and `in' modes on the right-hand wedge 
of the Penrose diagram representing the region exterior to the 
four-dimensional Schwarzschild black hole. The `up' modes are 
shown on the left-hand side and `in' modes on the right-hand side.}  
\label{fig:4DSchw:modes}
\end{center}
\end{figure}
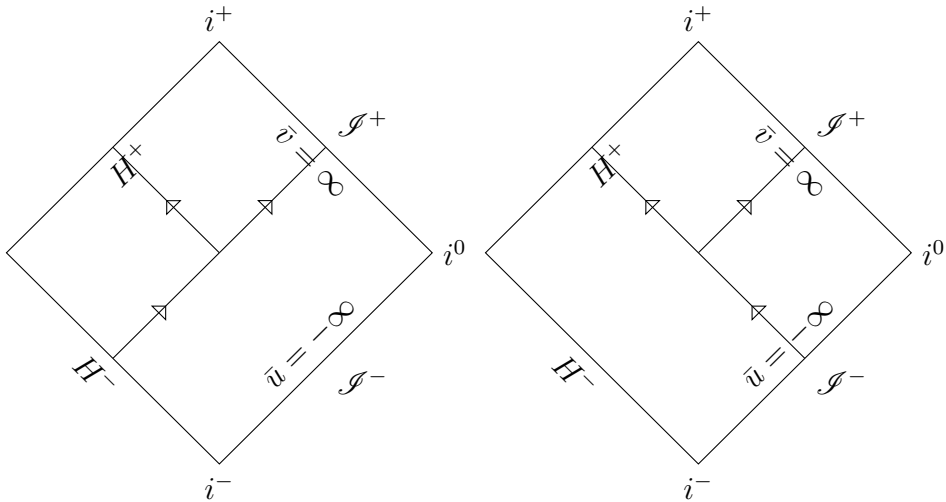

In the asymptotic limit of $r\to\infty$,
equation~\eqref{eq:4DSchw:radModPhi} has solutions $\phi(r)\approx
\expo^{\pm i\omega r*}/r$. The mode solutions with the simple form
$\expo^{+i\omega r*}/r$ as the leading order term at infinity are
known as `up-modes', and despite being of this simple outgoing form at
infinity they are a linear superposition of ingoing and outgoing modes
at the horizon. Conversely, we have mode solutions known as `in-modes'
that take on a simple ingoing form at the horizon, $\expo^{-i\omega
  r*}/r$, but because of scattering from the potential term
in~\eqref{eq:4DSchw:radModTort} they are a linear superposition of
ingoing and outgoing modes at infinity. The up- and in-modes are
illustrated schematically in Figure~\ref{fig:4DSchw:modes}.
%
%
%

\subsection{Numerical methods for obtaining the boundary conditions}
\label{sec:4DSchw:bcs}
\par Our first task is to find boundary conditions for both the in-modes and up-modes. With these boundary conditions for $\phin,\phup$ and $\phin^{\prime},\phup^{\prime}$, we can numerically solve the ODE~\eqref{eq:4DSchw:radModPhi} to high precision using the Mathematica (TM) function `NDSolve' .
\subsubsection{Boundary conditions for the up-modes}
The up-modes take on the simple form $\phup \sim \expo^{i\omega r^{*}}/r$ as $r\to\infty$, and they are illustrated on the left-hand side of Figure~\ref{fig:4DSchw:modes}. To numerically obtain their value at a given suitably large radius, which we denote by $\rinf$, we substitute the ansatz
\begin{equation}
\phup\sim\frac{\expo^{i\omega r^{*}}}{r}\expo^{v(r)}\, ,
\label{eq:4DSchw:upAnsatz}
\end{equation}
with
\begin{equation}
v(r):=\sum_{n=1}^{\infty}\frac{c_n}{r^n}\,,
\label{eq:4DSchw:v}
\end{equation}
into~\eqref{eq:4DSchw:radModPhi}. This leads to an equation for $v(r)$:
\begin{equation}
\begin{aligned}
&r^2(r-2M)v^{\prime\prime}(r)+r^2(r-2M)(v^{\prime}(r))^2+2r\left(M+i\omega r^2\right)v^{\prime}(r)\\
&\qquad\qquad\qquad\qquad\qquad\qquad\qquad-\left(\ell(\ell+1)r+2M\right)=0\, .
\label{eq:4DSchw:veq}
\end{aligned}
\end{equation} 
We substitute \eqref{eq:4DSchw:v} into \eqref{eq:4DSchw:veq} and collect inverse powers of~$r$. The coefficient of each power of $r$ must be set equal to zero. The lowest power leads to an equation only involving ~$c_1$, the next power only involves $c_1$ and~$c_2$, 
the next only $c_1$, $c_2$ and $c_3$, and so on. 
Starting with~$c_1$, we iteratively solve for the $c_i$ by substituting the previous result into the next equation to be solved. In practice, the upper limit in the sum \eqref{eq:4DSchw:v} is replaced by some suitable cut-off, denoted as~$\ninf$. 
This means that the highest power we can trust in the $r^{-1}$ expansion of \eqref{eq:4DSchw:veq} is $r^{-\left(\ninf-2\right)}$, 
and the highest coefficient obtained is~$c_{\ninf}$. The values of $\ninf$ and $\rinf$ are determined by the desired numerical accuracy.
\par The initial conditions for $\phup$ and $\phup^{\prime}$ are computed using these $c_i$ and by evaluating at $\rinf$:
\begin{equation}
\begin{aligned}
\phup(\rinf)&=\frac{\expo^{i\omega r^{*}(\rinf)}}{\rinf}\expo^{v(\rinf)}\,,\\
\phup^{\prime}(\rinf)&=\frac{d}{dr}\left[\frac{\expo^{i\omega r^{*}(r)}}{r}\expo^{v(r)}\right]_{r=\rinf}\, .
\label{eq:4DSchw:inits}
\end{aligned}
\end{equation}
These initial conditions become more accurate as $\rinf$ and $\ninf$ increase.
\par
We computed the initial conditions~\eqref{eq:4DSchw:inits} in Mathematica (TM) for $\ninf=100$ and $\rinf=15000M$, where we set $M=1$ in the code and re-inserted the appropriate factors of $M$ in the computed physical answers by dimensional analysis. 
Having computed the boundary conditions, we then used Mathematica's `NDSolve' function to generate our up-modes $\phup$ for a given $(\omega,\ell)$. 
We sought a result for the transition rate that was accurate to around 3 or 4 decimal places. As we shall see later, the Wightman function is constructed using tens of thousands of points in $(\omega,\ell)$ parameter space, and a high precision in the individual $\phup,\phin$ modes is essential. In order to get the desired accuracy results for the transition rate, we used the following precision settings in `NDSolve'; we set `WorkingPrecision' to around 40, `AccuracyGoal' to around 32 and `PrecisionGoal' to around 20. With these settings, the results for $\phup,\phin$ did not change to around 10 decimal places upon further increases to the `NDSolve' precision settings. 
\subsubsection{Boundary conditions for the in-modes}
The in-modes are the modes that take on a simple ingoing form at the horizon, $\expo^{-i\omega r^{*}}/r$, but at any finite radius are a complicated superposition of ingoing and outgoing plane waves because of the scattering from the gravitational potential. They are illustrated on the right-hand side of Figure~\ref{fig:4DSchw:modes}. Thus, our strategy is to compute the initial conditions of the in-modes at the horizon, taking
\begin{equation}
\phin\sim \frac{\expo^{-i\omega r^{*}}}{r}w(r)\, 
\label{eq:4DSchw:inAnsatz}
\end{equation} 
as our ansatz, with 
\begin{equation}
w(r):=\sum^{\infty}_{n=0} b_n \left(r-2M\right)^n\, ,
\label{eq:4DSchw:w}
\end{equation}
and $b_0=1$. 

We substitute~\eqref{eq:4DSchw:inAnsatz} into~\eqref{eq:4DSchw:radModPhi} to obtain an equation in $w(r)$ that reads
\begin{equation}
r^2 (r-2M)w^{\prime\prime}(r)+2r\left(M-ir^2\omega\right)w^{\prime}(r)-(\ell(\ell+1)r+2M)w(r)=0\, .
\label{eq:4DSchw:weq}
\end{equation} 
Using~\eqref{eq:4DSchw:w} in~\eqref{eq:4DSchw:weq}, a recursion relation can be obtained~\cite{leaver}:
\begin{equation}
\begin{aligned}
&b_0=1\,, b_{-1}=b_{-2}=0\,,\\
&b_n=-\frac{\left[-12i\omega M(n-1)+(2n-3)(n-1)-(\ell(\ell+1)+1)\right]}{2M\left(n^2-i4Mn\omega\right)}b_{n-1}\\
&-\frac{\left[(n-2)(n-3)-i12M\omega(n-2)-\ell(\ell+1)\right]}{4M^2\left(n^2-i4Mn\omega\right)}b_{n-2}\\
&+\frac{i\omega(n-3)}{2M^2\left(n^2-i4Mn\omega\right)}b_{n-3}\, .
\label{eq:4DSchw:recursion}
\end{aligned}
\end{equation}
We are now in a position to compute the initial conditions for $\phin$ and $\phin^{\prime}$. We use these $b_i$ with the upper limit of the sum~\eqref{eq:4DSchw:w} replaced by some finite integer $\nh$, determined by the accuracy requirements, and we evaluate at the near horizon radius $\rh$, obtaining
\begin{equation}
\begin{aligned}
\phin(\rh)&=\frac{\expo^{-i\omega r^{*}(\rh)}}{\rh}w(\rh)\,,\\
\phin^{\prime}(\rh)&=\frac{d}{dr}\left[\frac{\expo^{-i\omega r^{*}(r)}}{r}w(r)\right]_{r=\rh}\, .
\label{eq:4DSchw:HorizonInits}
\end{aligned}
\end{equation}
In practice, the initial conditions \eqref{eq:4DSchw:HorizonInits} were computed in Mathematica (TM) for $\nh=200$ and $\rh=(20,000,001/10,000,000)M$. 
Given these boundary conditions, we used Mathematica's 
`NDSolve' function to generate our in-modes $\phin$ for a given $(\omega,\ell)$, 
with the same precision settings as for the up-modes.
\subsection{Normalisation}
\label{sec:4DSchw:norm}
We choose a basis whose asymptotic behaviour as $r^{*}\to\pm\infty$ is
\begin{equation}
\tPhin(r)\sim 
\begin{cases}
B^{\text{in}}_{\omega\ell}\expo^{-i\omega r^{*}}, & \,\, r\to 2M \, , \\
r^{-1}\expo^{-i\omega r^{*}}+A^{\text{in}}_{\omega\ell}r^{-1}\expo^{+i\omega r^{*}},  & \,\, r\to \infty \, ,
\label{eq:4DSchw:asyIn}
\end{cases}
\end{equation}
and 
\begin{equation}
\tPhup(r)\sim 
\begin{cases}
A^{\text{up}}_{\omega\ell}\expo^{-i\omega r^{*}}+\expo^{+i\omega r^{*}}, & \,\, r\to 2M \, , \\
B^{\text{up}}_{\omega\ell}r^{-1}\expo^{+i\omega r^{*}},  & \,\, r\to \infty \, .
\label{eq:4DSchw:asyUp}
\end{cases}
\end{equation}
The reflection and transmission coefficients satisfy the following Wronskian relations:
\begin{equation}
\begin{aligned}
\Bup&=(2M)^2\Bin\,,\\   
|\Ain|^2&=1-4M^2|\Bin|^2\,,\\    
|\Ain|^2&=|\Aup|^2\,,\\
|\Aup|^2&=1-\frac{|\Bup|^2}{4M^2}\, ,  
\label{eq:4DSchw:Wronsk}
\end{aligned}
\end{equation}
and we can express the transmission and reflection coefficients by
\bea 
\Bup&=\frac{(2M)2i\omega}{W[\rin,\rup]}\,,\\
\Aup&=-\frac{W[\rin,\rup^{*}]^{*}}{W[\rin,\rup]}\,,\\
\Ain&=-\frac{W[\rin,\rup^{*}]}{W[\rin,\rup]}\,,
\label{eq:4DSchw:coeffsInCode}
\eea
where $\rin$ and $\rup$ are the unnormalised modes associated with equations~\eqref{eq:4DSchw:radModTort} and~\eqref{eq:4DSchw:radModPhi}, which are the modes we solve for in the Mathematica (TM) code.
\par It is convenient to replace $\tPhin$ and $\tPhup$ with the un-tilded modes $\Phin$ and $\Phup$, defined by
\begin{equation}
\begin{aligned}
\Phi^{\text{in}}_{\omega\ell}&=\tPhin\,,\\
\Phi^{\text{up}}_{\omega\ell}&=\frac{\tPhup}{2M}\, .
\end{aligned}
\end{equation}
Using the Wronskian relations~\eqref{eq:4DSchw:Wronsk}, it can be verified that the functions $R_{\omega,\ell}:=r\Phi_{\omega,\ell}$ are normalised as
\begin{equation}
\int^{\infty}_{-\infty}\,\mathrm{d}r^{*}\,R_{\omega_1,\ell}(r)R_{\omega_2,\ell}^{*}(r)=2\pi\delta(\omega_1-\omega_2)\,,
\label{eq:4DSchw:NormCond}
\end{equation}
where we have suppressed the superscripts ``in'' and ``up''.
\par The normalised modes in this basis can be expressed in terms of the modes that we explicitly solve for in Mathematica (TM), $\phup$ and $\phin$, which were discussed in Section~\ref{sec:4DSchw:bcs}. The result is
\begin{equation}
\begin{aligned}
\Phin &=\frac{\Bup}{2M}\phin(r)\,,\\
\Phup &=\frac{\Bup}{2M}\phup(r)\, .
\label{eq:4DSchw:relnAdrianModesNormModes}
\end{aligned}
\end{equation}
%
%
\par With this solution, we introduce the basis functions $u^{\text{in}}_{\omega\ell m}$ and $u^{\text{up}}_{\omega\ell m}$ by
\begin{equation}
\begin{aligned}
u^{\text{in}}_{\omega\ell m}(\x)&=\frac{1}{\sqrt{4\pi\omega}}\Phi^{\text{in}}_{\omega\ell}(r)Y_{\ell m}(\theta,\phi)\expo^{-i\omega t}\,,\\
u^{\text{up}}_{\omega\ell m}(\x)&=\frac{1}{\sqrt{4\pi\omega}}\Phi^{\text{up}}_{\omega\ell}(r)Y_{\ell m}(\theta,\phi)\expo^{-i\omega t}\, ,
\label{eq:4DSchw:BoulwareModes}
\end{aligned}
\end{equation}
where $\omega>0$. These modes are positive frequency with respect to the Schwarzschild time translation Killing vector $\partial_t$.
\par Using the Wronskian relations~\eqref{eq:4DSchw:Wronsk}, it can be verified that these modes satisfy the orthonormality relations
\bea 
\left(u^{\text{up}}_{\omega\ell m},u^{\text{up}}_{\omega^{\prime}\ell^{\prime} m^{\prime}}\right)&=\delta_{\ell \ell^{\prime}}\delta_{m m^{\prime}}\delta\left(\omega-\omega^{\prime}\right)\,,\\
\left(u^{\text{in}}_{\omega\ell m},u^{\text{in}}_{\omega^{\prime}\ell^{\prime} m^{\prime}}\right)&=\delta_{\ell \ell^{\prime}}\delta_{m m^{\prime}}
\delta\left(\omega-\omega^{\prime}\right)\,,\\
\left(u^{\text{in}}_{\omega\ell m},u^{\text{up}}_{\omega^{\prime}\ell^{\prime} m^{\prime}}\right)&=0\,,
\eea
where the Klein-Gordon (indefinite) inner product on a spacelike hyperplane of simultaneity at instant $t$ is defined by
\be 
( \phi, \chi)=-i\int_{2M}^{\infty}\,\mathrm{d}r\,\frac{\,r^2}{(1-2M/r)}\int^{\pi}_0\,\mathrm{d}\theta\,\sin{\theta}\,\int^{2\pi}_0\,\mathrm{d}\phi\, \left[\phi\partial_t \chi^{*}-\left(\partial_t \phi\right)\chi^{*}\right]\,.
\ee
The complex conjugate modes satisfy similar orthonormality relations with a minus sign, and the inner product relation between the modes~\eqref{eq:4DSchw:BoulwareModes} and the complex conjugates vanish.
\section{Which quantum state?}
\label{sec:WhichVac}
We shall analyse the transition rate when the field is in 
the Hartle-Hawking state, 
the Boulware state and the Unruh state. 
In this section, we provide a brief reminder 
of the properties of each of the three states.
\subsection{The Hartle-Hawking state}
\par 
The Hartle-Hawking state is regular across the both the past and future horizon, and it reduces to a thermal heat bath at spatial infinity with temperature $T_H=\kappa/(2\pi)$, where the surface gravity, $\kappa$, is defined by $\kappa:=1/(4M)$. 
\par In order to construct the Wightman function for the quantum field in the Hartle-Hawking state, we must expand the quantum field in terms of the modes that have the analytic properties of positive-frequency plane waves with respect to the horizon generators; these modes take the form~\cite{byd,cyf}
\begin{equation}
\begin{aligned}
w^{\text{in}}_{\omega\ell m}&=\frac{1}{\sqrt{2\sinh{\left(4\pi M\omega\right)}}}\left(\expo^{2\pi M\omega}u^{\text{in}}_{\omega\ell m}+\expo^{-2\pi M\omega}v^{\text{in}*}_{\omega\ell m}\right)\,,\\
\bar{w}^{\text{in}}_{\omega\ell m}&=\frac{1}{\sqrt{2\sinh{\left(4\pi M\omega\right)}}}\left(\expo^{-2\pi M\omega} u^{\text{in}*}_{\omega\ell m}+\expo^{2\pi M\omega}v^{\text{in}}_{\omega\ell m}\right)\,,\\
w^{\text{up}}_{\omega\ell m}&=\frac{1}{\sqrt{2\sinh{\left(4\pi M\omega\right)}}}\left(\expo^{2\pi M\omega}u^{\text{up}}_{\omega\ell m}+\expo^{-2\pi M\omega}v^{\text{up}*}_{\omega\ell m}\right)\,,\\
\bar{w}^{\text{up}}_{\omega\ell m}&=\frac{1}{\sqrt{2\sinh{\left(4\pi M\omega\right)}}}\left(\expo^{-2\pi M\omega} u^{\text{up}*}_{\omega\ell m}+\expo^{2\pi M\omega}v^{\text{up}}_{\omega\ell m}\right)\,,
\label{eq:4DSchw:UnruhModes}
\end{aligned}
\end{equation}
where the $v$ are functions analogous to $u$ on the second exterior region of the Kruskal manifold. The modes are extended to the full Kruskal manifold by analytic continuation.
\par 
Expanding the quantum field $\psi$ in terms of these modes
gives 
\begin{align}
\psi
&=\sum^{\infty}_{\ell=0}\sum^{+\ell}_{m=-\ell}\int^{\infty}_0\,\mathrm{d}\omega\,
\Bigl(d^{\text{up}}_{\omega\ell m}w^{\text{up}}_{\omega\ell m}
+\bar{d}^{\text{up}}_{\omega\ell m}\bar{w}^{\text{up}}_{\omega\ell m}
+d^{\text{in}}_{\omega\ell m}w^{\text{in}}_{\omega\ell m}
+\bar{d}^{\text{in}}_{\omega\ell m}\bar{w}^{\text{in}}_{\omega\ell
  m}\Bigr)
\notag 
\\[1ex]
& 
\hspace{24ex}
+\text{h.c.}\,.
\label{eq:4DSchw:FieldExp}
\end{align}
The $d^{\text{a}}$ and $\bar{d}^{\text{a}}$ ($d^{\text{a}\,\dag}$ and
$\bar{d}^{\text{a}\,\dag}$) operators, with $\text{a} \in
\{\text{in},\text{up}\}$, 
are the annihilation (creation) operators with respect 
to the $w$ and $\bar{w}$ modes, and they satisfy
\bea 
\left[d^{\text{a}}_{\omega\ell m},
d^{\text{a}'\,\dag}_{\omega^{\prime}\ell^{\prime} m^{\prime}}\right]
&=\delta\left(\omega-\omega^{\prime}\right)
\delta_{\text{a}\text{a}'}
\delta_{\ell\ell^{\prime}}
\delta_{m m^{\prime}}\,,\\
\left[\bar{d}^{\text{a}}_{\omega\ell m},
\bar{d}^{\text{a}'\,\dag}_{\omega^{\prime}\ell^{\prime} m^{\prime}}\right]
&=\delta\left(\omega-\omega^{\prime}\right)
\delta_{\text{a}\text{a}'}
\delta_{\ell \ell^{\prime}}
\delta_{m m^{\prime}}\,
\label{eq:4DSchw:commd}
\eea
with the commutators between barred and unbarred operators vanishing, 
and
\be
d^{\text{a}}_{\omega\ell m}|0_K\rangle =\bar{d}^{\text{a}}_{\omega\ell m}|0_K\rangle =0\,.
\ee
The state $|0_K \rangle$ is the Hartle-Hawking state, and it is normalised such that
\be 
\langle 0_K|0_K \rangle=1\,.
\ee
In the exterior region of the hole, the modes~\eqref{eq:4DSchw:UnruhModes} reduce to a simple form because the $v$ functions vanish, and if we compute the Wightman function for the Hartle-Hawking state in the exterior region, we find
\begin{equation}
\begin{aligned}
&W(\x,\x'):=\langle 0_K|\psi(\x)\psi(\x')|0_K\rangle\\
&=\sum^{\infty}_{\ell=0}\sum^{+\ell}_{m=-\ell}\int^{\infty}_0\,\mathrm{d}\omega\, \frac{1}{8\pi\omega\sinh{\left(4\pi M\omega\right)}}\times\\
&\Bigg[\expo^{4\pi M\omega-i\omega\Delta t}Y_{\ell m}(\theta,\phi)Y^{*}_{\ell m}(\theta',\phi')\left(\Phup(r)\Phup^{*}(r')+\Phin(r)\Phin^{*}(r')\right)\\
&+\expo^{-4\pi M\omega+i\omega\Delta t}Y^{*}_{\ell m}(\theta,\phi)Y_{\ell m}(\theta',\phi')\left(\Phup^{*}(r)\Phup(r')+\Phin^{*}(r)\Phin(r')\right)\Bigg]\,,
\label{eq:4DSchw:HHWightman}
\end{aligned}
\end{equation}
with $\Delta t:=t-t'$.
\subsection{The Boulware state}
\par The Boulware state is analogous to the Rindler state in Rindler spacetime, and it is not regular across the black hole horizon. The Boulware state reduces to the Minkowski vacuum at spatial infinity. To construct the Wightman function for the Boulware state, 
the quantum scalar field is expanded in terms of the modes~\eqref{eq:4DSchw:BoulwareModes}, i.e.
\begin{equation}
\begin{aligned}
\psi=\sum^{\infty}_{\ell=0}\sum^{+\ell}_{m=-\ell}\int^{\infty}_0\,\mathrm{d}\omega\,\Bigl(b^{\text{up}}_{\omega\ell m}u^{\text{up}}_{\omega\ell m}+b^{\text{in}}_{\omega\ell m}u^{\text{in}}_{\omega\ell m}\Bigr)+\text{h.c.}\,,
\end{aligned}
\end{equation}
where the $b$ and $b^{\dag}$ operators are 
respectively the annihilation 
and creation operators for the $u$ modes that 
satisfy the commutation relations
\be
\left[
b^{\text{a}}_{\omega\ell m},
b^{\text{a}'\,\dag}_{\omega^{\prime}\ell^{\prime} m^{\prime}} 
\right]
=\delta\left(\omega-\omega^{\prime}\right)
\delta_{\text{a}\text{a}'}
\delta_{\ell\ell^{\prime}}
\delta_{m m^{\prime}}\,
\label{eq:4DSchw:commb}
\ee
with $\text{a} \in \{\text{in},\text{up}\}$.
The Boulware state $|0_B \rangle$ is defined by 
\be
b^{\text{a}}_{\omega\ell m}|0_B\rangle =0 
\ee
and normalised such that 
\be 
\langle 0_B|0_B \rangle=1\,.
\ee
Hence, in the exterior region, the Wightman function of a scalar field in 
the Boulware state can be expressed as
\begin{align}
W(\x,\x')
& :=
\langle 0_B|\psi(\x)\psi(\x')|0_B\rangle
\notag
\\
&\ =\sum^{\infty}_{\ell=0}\sum^{+\ell}_{m=-\ell}\int^{\infty}_0\,\mathrm{d}\omega\,
\frac{Y_{\ell m}(\theta,\phi)
Y^{*}_{\ell m}(\theta',\phi')}{4\pi\omega}
\expo^{-i\omega\Delta t}
\notag 
\\
& \hspace{20ex}
\times 
\left(\Phup(r)\Phup^{*}(r')+\Phin(r)\Phin^{*}(r')\right)\,.
\label{eq:4DSchw:preBoulwareStaticW}
\end{align}
\subsection{The Unruh state}
The Unruh state mimics the geometric effects of a collapsing star, and
it represents a time-asymmetric flux of radiation from the black
hole. The Unruh mode construction \eqref{eq:4DSchw:UnruhModes} is
applied only to the up-modes that originate on $\scrh^{-}$ and not to
the in-modes originating on~$\scri^{-}$. Hence, the Wightman function
in the Unruh state is defined by first expanding the quantum scalar
field as
\begin{equation}
\begin{aligned}
\psi=\sum^{\infty}_{\ell=0}\sum^{+\ell}_{m=-\ell}\int^{\infty}_0\,\mathrm{d}\omega\,\Bigl(d^{\text{up}}_{\omega\ell m}w^{\text{up}}_{\omega\ell m}+\bar{d}^{\text{up}}_{\omega\ell m}\bar{w}^{\text{up}}_{\omega\ell m}+b^{\text{in}}_{\omega\ell m}u^{\text{in}}_{\omega\ell m}\Bigr)+\text{h.c.}\,,
\end{aligned}
\end{equation}
where now
\begin{equation}
b^{\text{in}}_{\omega\ell m}|0_U\rangle=d^{\text{up}}_{\omega\ell m}|0_U\rangle =\bar{d
}^{\text{up}}_{\omega\ell m}|0_U\rangle=0\,,
\end{equation}
with $|0_U \rangle$ the Unruh state. 
The annihilation and creation operators $b$,~$d$ and $b^{\dag}$,~$d^{\dag}$ satisfy the commutation relations given in~\eqref{eq:4DSchw:commd} and~\eqref{eq:4DSchw:commb}.
\par
Hence, the Wightman function of a scalar field in this state can be expressed as
\begin{equation}
\begin{aligned}
&W(\x,\x'):=\langle 0_U|\psi(\x)\psi(\x')|0_U\rangle\\
&=\sum^{\infty}_{\ell=0}\sum^{+\ell}_{m=-\ell}\int^{\infty}_0\,\mathrm{d}\omega\,\left[w^{\text{up}}_{\omega\ell m}(\x)w^{\text{up}*}_{\omega\ell m}(\x')+\bar{w}^{\text{up}}_{\omega\ell m}(\x)\bar{w}^{\text{up}*}_{\omega\ell m}(\x')+u^{\text{in}}_{\omega\ell m}(\x)u^{\text{in}*}_{\omega\ell m}(\x')\right]\,.
\end{aligned}
\end{equation}
In the exterior region, this reduces to
\begin{equation}
\begin{aligned}
W(\x,\x')=\sum^{\infty}_{\ell=0}\sum^{+\ell}_{m=-\ell}\int^{\infty}_0\,\mathrm{d}\omega & \Bigg[\frac{\expo^{4\pi M\omega-i\omega\Delta t}Y_{\ell m}(\theta,\phi)Y^{*}_{\ell m}(\theta',\phi')\Phup(r)\Phup^{*}(r')}{8\pi\omega\sinh{\left(4\pi M\omega\right)}}\\
&+\frac{\expo^{-4\pi M\omega+i\omega\Delta t}Y^{*}_{\ell m}(\theta,\phi)Y_{\ell m}(\theta',\phi')\Phup^{*}(r)\Phup(r')}{8\pi\omega\sinh{\left(4\pi M\omega\right)}}\\
&+\frac{\expo^{-i\omega\Delta t}Y_{\ell m}(\theta,\phi)Y^{*}_{\ell m}(\theta',\phi')\Phin(r)\Phin^{*}(r')}{4\pi\omega}\Bigg]\,,\\
\label{eq:4DSchw:UnruhWightman}
\end{aligned}
\end{equation}
with $\Delta t:=t-t'$.
\section{Static detector}
\label{sec:4DSchw:static}
In this section, we specialise to a static detector: $r=r'=R$, $\Delta t=\Delta\tau/\sqrt{1-2M/R}$, and we can take $\theta=\theta'=\phi=\phi'=0$ without loss of generality.

\subsection{Hartle-Hawking state}
\label{sec:static:HH}
When the detector is static, the Wightman function of the Hartle-Hawking state in the exterior region~\eqref{eq:4DSchw:HHWightman} reduces to the form 
\begin{equation}
\begin{aligned}
&W(\x,\x')=\sum_{\ell}\int^{\infty}_0\,\mathrm{d}\omega\, \frac{\left(2\ell+1\right)}{16\pi^2\omega\sinh{\left(4\pi M\omega\right)}}
\left(|\Phup(R)|^2+|\Phin(R)|^2\right)\\
&\qquad\qquad\qquad\qquad\times\cosh{\left[4\pi M\omega-\frac{i\omega \Delta\tau}{\sqrt{1-2M/R}}\right]}\,,
\label{eq:4DSchw:HHWightmanStatic}
\end{aligned}
\end{equation}
where we have used (14.30.4) from~\cite{dlmf} to collapse the $m$-sum.
\par We now substitute~\eqref{eq:4DSchw:HHWightmanStatic} into the expression for the transition rate~\eqref{eq:techIntro:transRate}.
After interchanging the order of the $s$- and $\omega$-integrals and taking the regulator to zero, we arrive at
\begin{equation}
\begin{aligned}
\mathcal{\dot{F}}\left(E\right)&=\int^{\infty}_0\,\mathrm{d}\omega\,\sum^{\infty}_{l=0}\frac{\left(2\ell+1\right)}{16\pi^2\omega\sinh{\left(4\pi M\omega\right)}}\left(|\Phup(R)|^2+|\Phin(R)|^2\right)\\
&\qquad\qquad\qquad\times\int^{\infty}_{-\infty}\,\mathrm{d}s\,\expo^{-iEs}\cosh{\left[4\pi M\omega-\frac{i\omega s}{\sqrt{1-2M/R}}\right]}\,.
\end{aligned}
\end{equation}
The $s$-integral can be computed analytically, resulting in
\begin{equation}
\begin{aligned}
&\mathcal{\dot{F}}\left(E\right)=\int^{\infty}_0\,\mathrm{d}\omega\,\sum^{\infty}_{l=0}\frac{\left(2\ell+1\right)}{16\pi\omega\sinh{\left(4\pi M\omega\right)}}\left(|\Phup(R)|^2+|\Phin(R)|^2\right)\\[1ex]
&\ \ \ 
\times\left[\expo^{4\pi M\omega}\delta\left(E+\frac{\omega}{\sqrt{1-2M/R}}\right)+\expo^{-4\pi M\omega}\delta\left(E-\frac{\omega}{\sqrt{1-2M/R}}\right)\right]\,.
\end{aligned}
\end{equation}
The factors $|\Phup(R)|$ and $|\Phin(R)|$ can be extended to negative values of $\omega$ by symmetry. This allows one to write the transition rate as
\begin{equation}
\begin{aligned}
&\mathcal{\dot{F}}\left(E\right)=\sum^{\infty}_{l=0}\frac{\left(2\ell+1\right)}{4\pi}\sqrt{1-2M/R}\left(|\Phi^{\text{up}}_{\tilde{\omega}\ell}(R)|^2+|\Phi^{\text{in}}_{\tilde{\omega}\ell}(R)|^2\right)\times\\
&\Bigg[\frac{\expo^{-4\pi ME \sqrt{1-2M/R}}\Theta(-E)}{-4E\sqrt{1-2M/R}\sinh{\bigl(-4\pi ME\sqrt{1-2M/R}\,\bigr)}}\\
&\quad\quad\quad\quad\quad\quad\quad\quad+\frac{\expo^{-4\pi ME \sqrt{1-2M/R}}\Theta(E)}{4E\sqrt{1-2M/R}\sinh{\bigl(4\pi ME\sqrt{1-2M/R}\bigr)}}\Bigg]\,,
\end{aligned}
\end{equation}
where $\tilde{\omega}:=E\sqrt{1-2M/R}$. This can further be simplified to
\begin{equation}
\begin{aligned}
&\mathcal{\dot{F}}\left(E\right)=\frac{1}{8\pi E}\frac{1}{\expo^{E/T_{\text{loc}}}-1}\sum^{\infty}_{l=0}\left(2\ell+1\right)\left(|\Phi^{\text{up}}_{\tilde{\omega}\ell}(R)|^2+|\Phi^{\text{in}}_{\tilde{\omega}\ell}(R)|^2\right)\, ,
\label{eq:4DSchw:staticHH_TR_result}
\end{aligned}
\end{equation}
where $T_{\text{loc}}$ is the local Hawking temperature, given by 
\begin{align}
T_{\text{loc}} := \frac{1}{8 \pi M \sqrt{1-2M/R}}
\,.
\label{eq:Tloc-def}
\end{align}
The non-Planckian factor in \eqref{eq:4DSchw:staticHH_TR_result} can
be thought of as the local density of states~\cite{Ottewill:1987tm}.
This result can be compared to the asymptotic form found
in~\cite{Candelas:1980zt}, but here we are most interested in
performing the calculation in the interesting region near the black
hole.
\par 
The result~\eqref{eq:4DSchw:staticHH_TR_result} manifestly obeys the
KMS condition by virtue of the fact that the modes
$\Phi^{\text{up}}_{\tilde{\omega}\ell}$ and
$\Phi^{\text{in}}_{\tilde{\omega}\ell}$ only depend on the absolute
value of $\tilde{\omega}:=E\sqrt{1-2M/R}$; hence, the modes only
depend on the absolute value of excitation energy. Thus, the condition
\begin{equation}
\mathcal{\dot{F}}\left(E\right)=
\expo^{-E/T_{\text{loc}}}\mathcal{\dot{F}}\left(-E\right)
\label{eq:4DSchw:KMS}
\end{equation}
is obeyed, and the transition rate is thermal in the
temperature~$T_{\text{loc}}$.  Our mode sum treatment, hence,
reproduces the thermality result that was deduced in
\cite{Hartle:1976tp} from the complex analytic properties of the
Wightman function.
\subsection{Boulware state}
\label{sec:static:Boul}
For the static detector and the field in the Boulware state, 
the Wightman function \eqref{eq:4DSchw:preBoulwareStaticW} reduces to
\begin{equation}
W(\x,\x')=\sum^{\infty}_{\ell=0}\int^{\infty}_0\,\mathrm{d}\omega\,\frac{\left(2\ell+1\right)}{16\pi^2\omega}\expo^{-i\omega\Delta\tau/\sqrt{1-2M/R}}\left(|\Phup(R)|^2+|\Phin(R)|^2\right)\,,
\label{eq:4DSchw:BoulwareStaticW}
\end{equation}
where again (14.30.4) in~\cite{dlmf} has been used.
\par We substitute the Wightman function~\eqref{eq:4DSchw:BoulwareStaticW} into transition rate~\eqref{eq:techIntro:transRate} and commute the $s$- and $\omega$-integrals to obtain
\begin{equation}
\begin{aligned}
\mathcal{\dot{F}}\left(E\right)=\int^{\infty}_0\,\mathrm{d}\omega\,\sum^{\infty}_{l=0}\frac{\left(2\ell+1\right)}{16\pi^2\omega}&\left(|\Phup(R)|^2+|\Phin(R)|^2\right)\\
&\times\int^{\infty}_{-\infty}\,\mathrm{d}s\,\expo^{-iEs}\expo^{-i\omega s/\sqrt{1-2M/R}}\,,
\end{aligned}
\end{equation}
and performing the $s$-integral gives
\begin{equation}
\begin{aligned}
\mathcal{\dot{F}}\left(E\right)=\int^{\infty}_0\,\mathrm{d}\omega\,\sum^{\infty}_{l=0}\frac{\left(2\ell+1\right)}{8\pi\omega}&\left(|\Phup(R)|^2+|\Phin(R)|^2\right)\\
&\times\delta\left(E+\frac{\omega}{\sqrt{1-2M/R}}\right)\, ,
\end{aligned}
\end{equation}
which can be simplified to
\begin{equation}
\begin{aligned}
\mathcal{\dot{F}}\left(E\right)=\frac{\Theta(-E)}{8\pi|E|}\sum^{\infty}_{l=0}\left(2\ell+1\right)\left(|\Phi^{\text{up}}_{\tilde{\omega}\ell}(R)|^2+|\Phi^{\text{in}}_{\tilde{\omega}\ell}(R)|^2\right)\,,
\label{eq:4DSchw:staticBoulware_TR_result}
\end{aligned}
\end{equation}
where $\tilde{\omega}:=E\sqrt{1-2M/R}$.
\par We note that when the field is in the Boulware state, the transition rate for the static detector is only non-zero for negative energies of the detector, i.e. de-excitations. The result~\eqref{eq:4DSchw:staticBoulware_TR_result} is very similar to the transition rate for the inertial detector in flat spacetime, $-E \Theta(-E)/2\pi$, only with modifications due to the curvature of spacetime. This is what one would expect for the Boulware state.
\subsection{Unruh state}
\label{sec:static:Unruh}
For the static detector and the field in the Unruh state, the Wightman function~\eqref{eq:4DSchw:UnruhWightman} reduces to
\begin{equation}
\begin{aligned}
&W(\x,\x')=\sum^{\infty}_{\ell=0}\int^{\infty}_0\,\mathrm{d}\omega\,\frac{\left(2\ell+1\right)}{16\pi^2\omega} \times\\
& \Bigg[\frac{|\Phup(R)|^2}{2\sinh{\left(4\pi M \omega\right)}}\left(\expo^{4\pi\omega-i\omega\Delta\tau/\sqrt{1-2M/R}}+\expo^{-4\pi\omega+i\omega\Delta\tau/\sqrt{1-2M/R}} \, \right)\\
&\,\,\,\,\,\,\,\quad\quad+|\Phin(R)|^2\expo^{-i\omega\Delta\tau/\sqrt{1-2M/R}}\Bigg]\,,
\label{eq:4DSchw:UnruhStaticW}
\end{aligned}
\end{equation}
where again (14.30.4) in~\cite{dlmf} has been used.
\par We substitute the Wightman function~\eqref{eq:4DSchw:UnruhStaticW} into transition rate~\eqref{eq:techIntro:transRate}, and after commuting the $\omega$- and $s$-integrals, we can compute the $s$-integrals analytically, as in the Hartle-Hawking and Boulware states static calculations. The result for the transition rate is
\begin{equation}
\mathcal{\dot{F}}\left(E\right)=\sum^{\infty}_{l=0}\frac{\left(2\ell+1\right)}{4\pi}\Bigg[\frac{|\Phi^{\text{up}}_{\tilde{\omega}\ell}(R)|^2}{2E\left(\expo^{E/T_{\text{loc}}}-1\right)}-\frac{|\Phi^{\text{in}}_{\tilde{\omega}\ell}(R)|^2}{2E}\Theta(-E)
\Bigg]\,,
\label{eq:4DSchw:staticUnruh_TR_result}
\end{equation}
where $\tilde{\omega}:=E\sqrt{1-2M/R}$ and $T_{\text{loc}}$ 
is given by~\eqref{eq:Tloc-def}. 
\section{Circular-geodesic detector}
\label{sec:4DSchw:circ}
In this section, we investigate the transition rate of a detector orbiting the Schwarzschild black hole on a circular geodesic. Explicitly, the detector trajectory is
\begin{equation}
r=R\,,\quad\theta=\pi/2\,,\quad t=a\tau\,,\quad \phi = a\Omega \tau\,,
\label{eq:4DSchw:circ:traj}
\end{equation}
where $R>3M$ and 
\begin{equation}
\begin{aligned}
&a:=\sqrt{R/(R-3M)}\,,\\
&\Omega:=\frac{d\phi}{dt}=\sqrt{M/R^3}\,.
\end{aligned}
\label{eq:4DSchw:circ:traj2}
\end{equation}
%
Trajectories with $3M<R\leq 6M$ are unstable, and trajectories with $R>6M$ are stable.
\subsection{Hartle-Hawking state}
\label{sec:circ:HH}
We first obtain the Wightman function for a detector on a circular geodesic in the Hartle-Hawking state by substituting~\eqref{eq:4DSchw:circ:traj} into \eqref{eq:4DSchw:HHWightman} and expanding the spherical harmonics. We obtain
\begin{equation}
\begin{aligned}
&W(\x,\x')=\sum^{\infty}_{\ell=0}\sum^{+\ell}_{m=-\ell}\int^{\infty}_0\,\mathrm{d}\omega\, \frac{(\ell-m)!(2\ell+1)|P^m_{\ell}(0)|^2}{32\pi^2\omega(l+m)!\sinh{\left(4\pi M\omega\right)}}\times\\
&\left(|\Phup(R)|^2+|\Phin(R)|^2\right)\left[\expo^{4\pi M\omega-ia \omega s+im a \Omega s}+\expo^{-4\pi M\omega+i a\omega s-im a \Omega s}\right]\,.
\label{eq:4DSchw:circ:W}
\end{aligned}
\end{equation}
Additionally, one can use (14.30.5) of~\cite{dlmf} to see that the contribution to the Wightman function will vanish unless $\ell+m$ is even. This means that for a given $\ell$ we can set $m \equiv \ell \,(\text{mod} 2)$. 
\par
We use~\eqref{eq:4DSchw:circ:W} in~\eqref{eq:techIntro:transRate}, and as in the static section, we can evaluate the $s$-integral analytically. The resulting expression reads
\begin{equation}
\begin{aligned}
&\mathcal{\dot{F}}\left(E\right)=\sum^{\infty}_{\ell=0}\sum^{\ell}_{m=-\ell}\int^{\infty}_0\,\mathrm{d}\omega\, \frac{(\ell-m)!(2\ell+1)|P^m_{\ell}(0)|^2}{16\pi\omega(l+m)!\sinh{\left(4\pi M\omega\right)}} 
\left(|\Phup(R)|^2+|\Phin(R)|^2\right)
\\[1ex]
&
\qquad\qquad\qquad
\times 
\left[\expo^{4\pi M\omega}\delta\left(E+a\omega-ma\Omega\right)+\expo^{-4\pi M\omega}\delta\left(E-a\omega+ma\Omega\right)\right]\,.
\label{eq:4DSchw:circ:HHrate1}
\end{aligned}
\end{equation}
Evaluating the integral over $\omega$, we finally obtain
\begin{equation}
\begin{aligned}
&\mathcal{\dot{F}}\left(E\right)=\sum^{\infty}_{\ell=0}\sum^{+\ell}_{m=-\ell} \frac{(\ell-m)!(2\ell+1)|P^m_{\ell}(0)|^2}{16\pi(l+m)!}\times\\
&\Bigg[\frac{\left(|\Phi^{\text{up}}_{\omega_{-}\ell}(R)|^2+|\Phi^{\text{in}}_{\omega_{-}\ell}(R)|^2\right)\expo^{4\pi M\omega_-}}{a\omega_{-}\sinh{\left(4\pi M\omega_{-}\right)}}\Theta(m a\Omega-E)\\
&\quad\quad\quad\quad+\frac{\left(|\Phi^{\text{up}}_{\omega_{+}\ell}(R)|^2+|\Phi^{\text{in}}_{\omega_{+}\ell}(R)|^2\right)\expo^{-4\pi M\omega_-}}{a\omega_{+}\sinh{\left(4\pi M\omega_{+}\right)}}\Theta(ma\Omega+E)\Bigg]\, ,
\label{eq:4DSchw:circ:HHrateRes}
\end{aligned}
\end{equation}
with
\begin{equation}
\begin{aligned}
\omega_{\pm}:=(ma\Omega\pm E)/a\,.
\end{aligned}
\end{equation}
\subsection{Boulware state}
\label{sec:circ:Boul}
We start by substituting~\eqref{eq:4DSchw:circ:traj} into~\eqref{eq:4DSchw:preBoulwareStaticW}, and we expand the spherical harmonics. The Wightman function then reads
\begin{equation}
\begin{aligned}
W(\x,\x')&=\sum^{\infty}_{\ell=0}\sum^{\ell}_{m=-\ell}\int^{\infty}_0\,\mathrm{d}\omega\,\frac{(\ell-m)!(2\ell+1)|P^{m}_{\ell}(0)|^2}{16\pi^2\omega(\ell+m)!}\expo^{i ma\Omega\Delta\tau-ia\omega\Delta\tau}\\
&\qquad\qquad\qquad\qquad\times\left(|\Phup(R)|^2+|\Phin(R)|^2\right)\,.
\end{aligned}
\end{equation}
We substitute this Wightman function into~\eqref{eq:techIntro:transRate}, and we evaluate the $s$-integral analytically. The resulting expression for the transition rate is
\begin{equation}
\begin{aligned}
\mathcal{\dot{F}}\left(E\right)&=\sum^{\infty}_{\ell=0}\sum^{\ell}_{m=-\ell}\int^{\infty}_0\,\mathrm{d}\omega\,\frac{(\ell-m)!(2\ell+1)|P^{m}_{\ell}(0)|^2}{8\pi\omega(\ell+m)!}\left(|\Phup(R)|^2+|\Phin(R)|^2\right)\\
&\qquad\qquad\qquad\qquad\qquad\times\delta\left(a\omega-(ma\Omega-E)\right)\,.
\end{aligned}
\end{equation}
Evaluating the $\omega$-integral yields
\begin{equation}
\begin{aligned}
\mathcal{\dot{F}}\left(E\right)&=\frac{1}{a}\sum^{\infty}_{\ell=0}\sum^{\ell}_{m=-\ell}\frac{(\ell-m)!(2\ell+1)|P^{m}_{\ell}(0)|^2}{8\pi\omega_{-}(\ell+m)!}\left(|\Phi^{\text{up}}_{\omega_{-}\ell}(R)|^2+|\Phi^{\text{in}}_{\omega_{-}\ell}(R)|^2\right)\\
&\qquad\qquad\qquad\qquad\qquad\qquad\times\Theta\left(ma\Omega-E\right)\,,
\label{eq:4DSchw:circ:BoulwarerateRes}
\end{aligned}
\end{equation}
with
\begin{equation}
\omega_{-}:=(ma\Omega- E)/a\,.
\end{equation}
\subsection{Unruh state}
\label{sec:circ:Unruh}
This time we substitute~\eqref{eq:4DSchw:circ:traj} into \eqref{eq:4DSchw:UnruhWightman}, and we expand the spherical harmonics. The Wightman function then reads
\begin{align}
W(\x,\x')
&=
\sum^{\infty}_{\ell=0}\sum^{\ell}_{m=-\ell}\int^{\infty}_0\,
\mathrm{d}\omega\,\frac{(\ell-m)!(2\ell+1)|P^{m}_{\ell}(0)|^2}{16\pi^2(\ell+m)!}
\times
\notag
\\[1ex]
&\qquad\times 
\Bigg[\frac{|\Phup(R)|^2\left(\expo^{4\pi M\omega-ia\omega\Delta\tau+ima\Omega\Delta\tau}+\expo^{-4\pi M\omega+ia\omega\Delta\tau-ima\Omega\Delta\tau}\right)}{2\omega\sinh{\left(4\pi M\omega\right)}}
\notag
\\[1ex]
&\qquad\qquad\qquad\qquad\qquad\qquad
+\frac{|\Phin(R)|^2\expo^{-ia\omega\Delta\tau+ima\Omega\Delta\tau}}{\omega}\Bigg]\,.
\end{align}
Substituting this Wightman function into \eqref{eq:techIntro:transRate} and evaluating the $s$-integral analytically, the transition rate is
\begin{align}
&\mathcal{\dot{F}}\left(E\right)=\sum^{\infty}_{\ell=0}\sum^{\ell}_{m=-\ell}\int^{\infty}_0\,\mathrm{d}\omega\,\frac{(\ell-m)!(2\ell+1)|P^{m}_{\ell}(0)|^2}{8\pi(\ell+m)!}\times
\notag
\\
&\Bigg[\frac{|\Phup(R)|^2}{2\omega\sinh{\left(4\pi M\omega\right)}}\left(\expo^{4\pi M\omega}\delta\left(E+a\omega-ma\Omega\right)+\expo^{-4\pi M\omega}\delta\left(E-a\omega+ma\Omega\right)\right)
\notag
\\
&\quad\qquad\quad+\frac{|\Phin(R)|^2}{\omega}\delta\left(E+a\omega-ma\Omega\right)\Bigg]\,.
\end{align}
Evaluating the $\omega$-integral yields
\begin{align}
&\mathcal{\dot{F}}\left(E\right)=\frac{1}{a}\sum^{\infty}_{\ell=0}\sum^{\ell}_{m=-\ell}\frac{(\ell-m)!(2\ell+1)|P^{m}_{\ell}(0)|^2}{8\pi(\ell+m)!}\times
\notag
\\
&\Bigg[\left(\frac{|\Phi^{\text{up}}_{\omega_{-}\ell}(R)|^2}{2\omega_{-}\sinh{\left(4\pi M\omega_{-}\right)}}\expo^{4\pi M\omega_{-}}+\frac{|\Phi^{\text{in}}_{\omega_{-}\ell}(R)|^2}{\omega_{-}}\right)\Theta(ma\Omega-E)
\notag\\
&\quad\quad\quad\quad\quad\quad\quad\quad+\frac{|\Phi^{\text{up}}_{\omega_{+}\ell}(R)|^2}{2\omega_{+}\sinh{\left(4\pi M\omega_{+}\right)}}\expo^{-4\pi M\omega_{+}}\Theta(ma\Omega+E)\Bigg]\,,
\label{eq:4DSchw:circ:UnruhrateRes}
\end{align}
with
\begin{equation}
\omega_{\pm}:=(ma\Omega\pm E)/a\,.
\end{equation}
\subsection{Evaluation}
\label{sec:circ:eval}
\par It proves only necessary to compute $\Phi^{\text{up}}_{\omega_{\pm},\ell},\Phi^{\text{in}}_{\omega_{\pm},\ell}$, where $\omega_{\pm}:=(m a \Omega \pm E)/a$,  over the positive range $E>0,~m\geq 0$ in order to have all the data we need to reconstruct the full transition rate over both negative and positive $E$ and $m$. The reason for this is the fact that the absolute square of the modes only depends on the absolute value of $\omega$, and $\omega_{\pm}(m,E)$ can always be related to $\pm\omega_{\pm}(|m|,|E|)$. For example, assuming we wished to compute the $|\Phi^{\text{up}}_{\omega_{+},\ell}|^2,|\Phi^{\text{in}}_{\omega_{+},\ell}|^2$ for a term in the sum where both $E, m <0$, we can observe that
\bea
\omega_{+}(-|m|, -|E|)&=\frac{-|m|a\Omega-|E|}{a}\\
&=-\frac{|m|a\Omega+|E|}{a}\\
&=-\omega_{+}(|m|, |E|)\, .
\eea
Thus, if we have already computed the modes at $\omega_{+}(|m|, |E|)$, then by the fact that $|\omega_{+}(-|m|, -|E|)|=|\omega_{+}(|m|, |E|)|$ and the independence of $|\Phi^{\text{up}}_{\omega_{+},\ell}|^2,|\Phi^{\text{in}}_{\omega_{+},\ell}|^2$ on the overall sign of $\omega$, we see that we also have the value of the absolute value squared of the modes over the range where both $E,~m <0$. Further relations are 
\bea
\omega_{+}(-|m|,|E|)&=-\omega_{-}(|m|,|E|)\,,\\
\omega_{+}(|m|,-|E|)&=\omega_{-}(|m|,|E|)\,,\\
\omega_{-}(-|m|,-|E|)&=-\omega_{-}(|m|,|E|)\,,\\
\omega_{-}(-|m|,|E|)&=-\omega_{+}(|m|,|E|)\,,\\
\omega_{-}(|m|,-|E|)&=\omega_{+}(|m|,|E|)\,.
\eea
\section{Comparison with a Rindler observer}
\label{sec:4DSchw:compRind}
The analogy between the right-hand Rindler wedge and the exterior
Schwarzschild spacetime is well known~\cite{byd}. It seems a natural
question to ask whether the experience of the static detector when the
field is in the Hartle-Hawking state, which we have described in the
previous sections, is related to the experience of a detector in
Rindler spacetime on a Rindler trajectory with the field in the
Minkowski vacuum. Similarly, we ask if the experience of a detector on
a circular geodesic in Schwarzschild spacetime is related to that of a
detector on a Rindler trajectory but given some boost in the
transverse direction~\cite{Abdolrahimi:2013tha}.
\subsection{Static detector comparison with Rindler detector}
\par 
The Rindler observer's trajectory in $(3+1)$-dimensional Minkowski
spacetime is
\begin{equation}
\x(\tau)=\frac{1}{a}\bigl(\sinh{\left(a\tau\right)},\cosh{\left(a\tau\right)},L,0\bigr)\,,
\label{eq:4DSchw:circ:rindTraj}
\end{equation}
where the positive constant $a$ is the proper acceleration, $\tau$ is
the proper time, and we have introduced the real-valued constant $L$
for later convenience. 
With the quantum field in the Minkowski
vacuum, the transition rate for a detector on the Rindler
trajectory is \cite{byd}
\begin{equation}
\begin{aligned}
\mathcal{\dot{F}}\left(E\right)
=\frac{E}{2\pi\bigl(\expo^{2\pi E/a}-1\bigr)}\,,
\label{eq:4DSchw:circ:rindRate}
\end{aligned}
\end{equation}
which is thermal at the temperature~$a/(2\pi)$. 
We choose to compare the Rindler response to the Schwarzschild
response by matching the trajectories so that the Rindler temperature
$a/(2\pi)$ is equal to the local Hawking 
temperature $T_{\text{loc}}$~\eqref{eq:Tloc-def}. This gives 
\begin{equation}
a=1/\bigl(4M\sqrt{1-2M/R}\,\bigr)\,. 
\label{eq:4DSchw:circ:propAccel}
\end{equation}

We note that the proper acceleration of a static worldline in
Schwarzschild is given by
\begin{align}
a_S=M/\bigl(R^2\sqrt{1-2M/R}\,\bigr)
\,. 
\label{eq:schw-proper-acc-natural}
\end{align}
From \eqref{eq:4DSchw:circ:propAccel} and
\eqref{eq:schw-proper-acc-natural} we hence see that matching the
local temperatures is not the same as matching the proper
accelerations, although the two become asymptotically equal in the
near-horizon limit, where the analogy between the Rindler and
Schwarzschild spacetimes is the closest: as $R\to2M$, we
have $a\to\infty$ and $a_S\to\infty$ so that $a/a_S \to1$ and $a - a_S
\to0$.

\subsection{Circular-geodesic detector compared with 
Rindler plus transverse drift detector}
\par Next, consider the Rindler observer but with constant drift-velocity in the transverse $y$-direction:
\begin{equation}
\x(\tau')_{\text{drift}}=\frac{1}{a}
\bigl(\sinh{\left(q\tau'\right)},\cosh{\left(q\tau'\right)},p\tau',0\bigr)\,,
\label{eq:4DSchw:circ:driftTraj}
\end{equation}
where $a$, $q$ and $p$ are positive constants and $\tau'$ is the
proper time. 
In order for the four-velocity to 
be correctly normalised, we require that
\begin{equation}
a^2=q^2-p^2\,.
\label{eq:4DSchw:circ:paqRel}
\end{equation}
If we take $p\to 0$, this trajectory becomes the Rindler trajectory with proper acceleration $a$.
\par In Schwarzschild spacetime, the static detector has four-velocity given by
\be 
\U_{\text{static}}=\left(\sqrt{\frac{R}{R-2M}},0,0,0\right)\, ,
\ee
and the circular-geodesic trajectory, specified by 
\eqref{eq:4DSchw:circ:traj} and~\eqref{eq:4DSchw:circ:traj2}, has four-velocity 
\be
\U_{\text{circ}}=\left(\sqrt{\frac{R}{R-3M}},0,0,\sqrt{\frac{M}{R^2(R-3M)}}\right)\, .
\ee
It follows that
\begin{equation}
\U_{\text{circ}}\cdot\U_{\text{static}}=-\sqrt{\frac{R-2M}{R-3M}}\,.
\label{eq:4DSchw:circ:dotProd4velSchw}
\end{equation}

We choose to compare the Rindler detector with transverse drift (RDTD)
to the circular-geodesic Schwarzschild detector by matching the drift
velocity in Rindler to the orbital velocity in Schwarzschild. In
terms of the four-velocity vectors, this amounts to setting
$\U_{\text{RDTD}}\cdot\U_{\text{Rind}} =
\U_{\text{circ}}\cdot\U_{\text{static}}$, where $\U_{\text{Rind}}$ and
$\U_{\text{RDTD}}$ are the four-velocity of the Rindler detector and
RDTD respectively. As the circular geodesics exist only for $R>3M$, we
note that this comparison cannot be extended to the near-horizon
limit, where the analogy between the Rindler and Schwarzschild
spacetime is the closest.

To implement the matching, the dot product in 
$\U_{\text{RDTD}}\cdot\U_{\text{Rind}}$ must be evaluated when 
the Rindler and RDTD observers are at the same spacetime
point. Comparison of~\eqref{eq:4DSchw:circ:rindTraj}
and~\eqref{eq:4DSchw:circ:driftTraj} shows that in order to be at the
same point we must take $a\tau=q\tau'$ and $\tau'=L/p$.
This means that at this spacetime point
\begin{equation}
\begin{aligned}
&\U_{\text{Rind}}=\bigl(\cosh{\left(qL/p\right)},\sinh{\left(qL/p\right)},0,0\bigr)\,,\\
&\U_{\text{RDTD}}=\left(\frac{q}{a}\cosh{\left(qL/p\right)},\frac{q}{a}\sinh{\left(qL/p\right)},\frac{p}{a},0\right)\,,
\end{aligned}
\end{equation}
so that
\begin{equation}
\U_{\text{Rind}}\cdot\U_{\text{RDTD}}=-\frac{q}{a}\,.
\label{eq:4DSchw:circ:dotProd4velFlat}
\end{equation}
We want
\begin{equation}
q=\frac{1}{4M}\sqrt{\frac{R}{R-3M}}\,,
\label{eq:4DSchw:circ:q}
\end{equation}
and by virtue of~\eqref{eq:4DSchw:circ:propAccel}
and~\eqref{eq:4DSchw:circ:paqRel}, we have
\begin{equation}
p=\frac{1}{4M}\sqrt{\frac{MR}{(R-3M)(R-2M)}}\,.
\label{eq:4DSchw:circ:p}
\end{equation}
The transition rate for the RDTD can now easily be computed.
By~\eqref{eq:4DSchw:circ:driftTraj}, we first note that the Minkowski
interval is
\begin{equation}
\Delta\x^2=\frac{p^2}{a^2}\Delta\tau^2
-\frac{4}{a^2}\sinh^2{\left(\frac{q\Delta\tau}{2}\right)}\,.
\label{eq:4DSchw:circ:interval}
\end{equation}
This can be substituted into the transition rate found
in~\cite{louko-satz:profile}. The comparison will be examined in
Section~\ref{sec:4DSchw:results}. 

We also would like to see if the comparison between the detector on a
circular geodesic in Schwarzschild and the RDTD becomes better if we
make the transverse direction, in which the Rindler detector is
drifting, periodic. The proper-time period for the circular-geodesic
detector in Schwarzschild to complete a loop is
\begin{equation}
P:=2\pi\sqrt{R^2(R-3M)/M}\,.
\label{eq:4DSchw:circ:period}
\end{equation}
We wish to identify the transverse direction of Minkowski 
spacetime that our RDTD exists on by the same period in proper time. 
This means identifying the points
\begin{equation}
\begin{aligned}
y(\tau) &\sim y(\tau+nP)\\
&= y(\tau)+npP/a\,,
\end{aligned}
\end{equation}
where $n$ is an integer and $y$ is the transverse direction in which the drift occurs.
In order to get the transition rate of the RDTD on flat spacetime with periodic boundary conditions in the transverse drift direction, we employ the method of images. This results in the square interval
\begin{equation}
\Delta\x^2_n=-\frac{4}{a^2}\sinh^2{\left(\frac{q\Delta\tau}{2}\right)}
+\left(\frac{p\Delta\tau}{a}+\frac{npP}{a}\right)^2\,,\quad n \in \BbbZ\,.
\label{eq:4DSchw:circ:methImInterval}
\end{equation}
We substitute this interval into the transition rate~\eqref{eq:techIntro:transRate}. Because the periodicity could lead to singularities at $\Delta\tau\neq 0$, not dealt with by the Hadamard short distance form, we need the form of the transition rate with regulator intact. The exception, of course, is the $n=0$ term for which we can use the form of the transition rate found in~\cite{louko-satz:profile} with the regulator already taken to zero, see also~\cite{Hodgkinson:2012mr}, where such singularities were also encountered and dealt with.
For the $n\neq 0$ terms, the transition rate can be written as
\begin{align}
\mathcal{\dot{F}}\left(E\right)
&=
-\frac{a^2}{2q}\sum^{\infty}_{n=-\infty}\int^{\infty}_{-\infty}\,\frac{\mathrm{d}r\, \expo^{-2iEr/q}}
{\sinh^2 r-{\bigl(\frac{rp}{q}+\frac{npP}{2}\bigr)}^2}
\notag
\\[1ex]
&=-\frac{a^2}{4q}\sum^{\infty}_{n=-\infty}\int^{\infty}_{-\infty}\, 
\frac{\mathrm{d}r\,\expo^{-2iEr/q}}{\frac{rp}{q}+\frac{npP}{2}}
\notag
\\[1ex]
&\qquad\qquad\times
\left(\frac{1}{\sinh r-\bigl(\frac{rp}{q}+\frac{npP}{2}\bigr)}
-\frac{1}{\sinh r+\bigl(\frac{rp}{q}+\frac{npP}{2}\bigr)}\right)\,,
\label{eq:4DSchw:circ:n_not0_rate}
\end{align}
where the $i\epsilon$ prescription amounts to giving $r$ a small, negative, 
imaginary part near the singularities on the real axis.
\par We evaluate \eqref{eq:4DSchw:circ:n_not0_rate} numerically. 
We first use Mathematica's `FindRoot' function to solve the 
transcendental equations that specify the singularities in the 
integrand. With the singularities known, we compute the integral 
in \eqref{eq:4DSchw:circ:n_not0_rate} by using 
Mathematica's `CauchyPrincipalValue' method of `NIntegrate' 
and adding the contribution from the small semi-circle contours 
that pass around the singularities in the lower half-plane. 
The sum is cut off at some suitable value of $|n|$ 
when convergence has occurred to the desired precision.
\section{Results}
\label{sec:4DSchw:results}
\subsection{Static detector}
\label{sec:4DSchw:results:stat}
First, we look at the numerical results for the transition rate of 
a static detector at fixed radius $R$. 
We use the results \eqref{eq:4DSchw:staticHH_TR_result}, 
\eqref{eq:4DSchw:staticBoulware_TR_result} 
and \eqref{eq:4DSchw:staticUnruh_TR_result} 
to numerically obtain the transition rates in the 
Hartle-Hawking, Boulware and Unruh states respectively. 

We imposed a suitable cut-off in the $\ell$-sum that increased
with excitation energy (through $\tilde{\omega}$) and also increased
with increasing radius, $R$. Considering $R=4M$, for example, we
evaluated the transition rate at the points
$ME=-150/100,~-148/100,~\ldots,~148/100,~150/100$, excluding the $E=0$
point. The point $E=0$ is problematic because it would involve solving
for the modes at $\omega=0$, which proves difficult numerically. For
$R=4M$ and $M|E|=1/100$, we cut off the $\ell$-sum at $\ell=12$,
whereas at $M|E|=150/100$ we cut off the sum at $\ell=59$ (one could have
used much lower cut-off values quite adequately here, but in the
static case computation is fast and we could afford to use a larger
value for the cut-off than strictly necessary). For $R=40M$, we found
that at $M|E|=2$ a cut-off of $\ell=107$ was more than adequate as these
contributions had become negligibly small.

A final point to note is that because the differential equation 
\eqref{eq:4DSchw:radModPhi} depends on $\omega$ only via $\omega^2$, 
and in the static case we evaluate the modes at 
$\tilde{\omega}=E\sqrt{1-2M/R}$, the values of the modes
$|\phi^{\text{in}}_{\tilde{\omega}\ell}|^2$ 
and 
$|\phi^{\text{up}}_{\tilde{\omega}\ell}|^2$ 
only depend on $E$ through~$|E|$. Hence, we can just
evaluate over the positive range: $ME=2/100,~4/100,...,~150/100$, and then we immediately
have the values of
$|\phi^{\text{in}}_{\tilde{\omega}\ell}|^2$ and 
$|\phi^{\text{up}}_{\tilde{\omega}\ell}|^2$ over
the corresponding negative energies too.

\begin{figure}[t]  
\centering
\subfloat[$R=4M$]{\label{fig:static_R4M_3vacs}\includegraphics[width=0.48\textwidth]{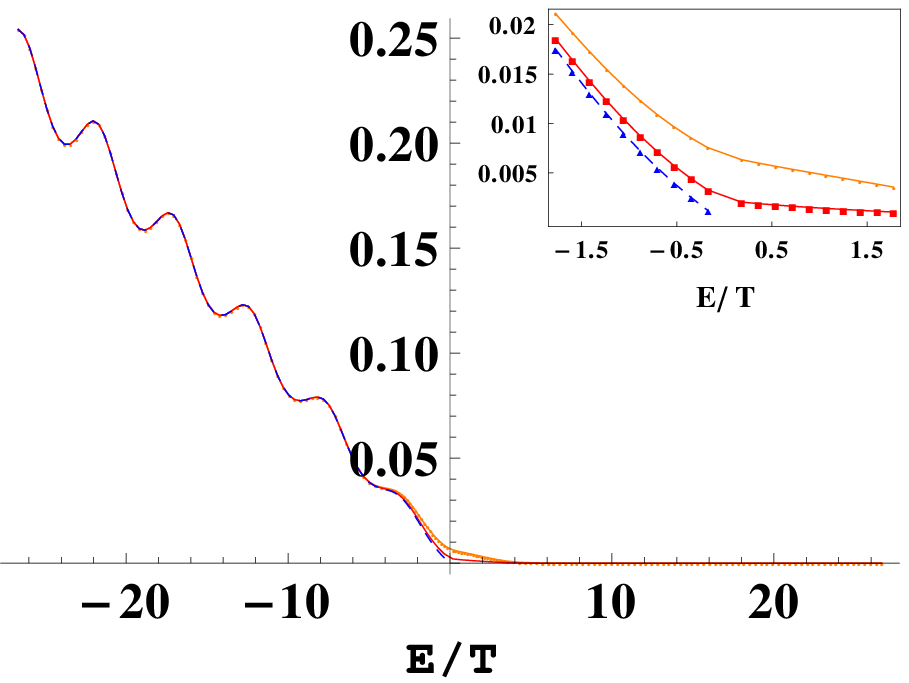}}\hspace{2pt}         
\subfloat[$R=40M$]{\label{fig:static_R40M_3vacs}\includegraphics[width=0.48\textwidth]{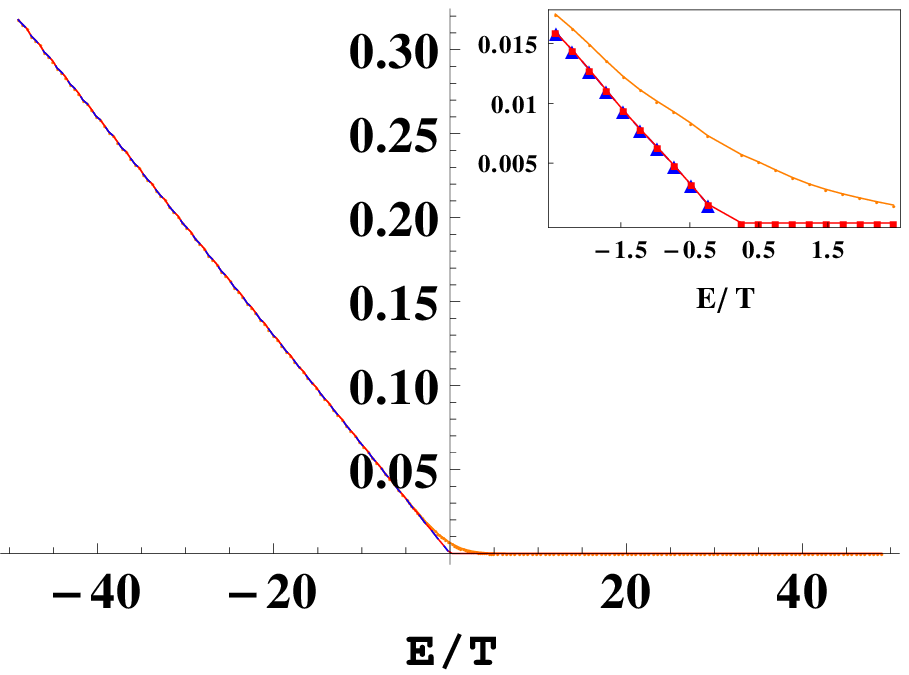}}  
\caption{$M\dot{\mathcal{F}}$ as a function of $E/T_{\text{loc}}$ for the static detector at various radii, showing the results for the Hartle-Hawking state (orange circles) computed from~\eqref{eq:4DSchw:staticHH_TR_result}, Boulware state (blue dashed) computed from~\eqref{eq:4DSchw:staticBoulware_TR_result} and Unruh state (red solid) computed from~\eqref{eq:4DSchw:staticUnruh_TR_result}.}
\label{fig:transrate_static_3vacs}
\end{figure}
\par 
Figure~\ref{fig:transrate_static_3vacs} shows the transition rate against 
the excitation energy of the detector divided by 
the local temperature $T_{\text{loc}}$~\eqref{eq:Tloc-def}. 
The horizon is at $R=2M$, and we see that as we move away from the horizon, 
far from the hole at $R=40M$, the transition rates for the Boulware and Unruh 
states align uniformly across negative energy gap. 
Near the horizon, at $R=4M$, the transition rate is seen to oscillate for large, 
negative energy gap. 
Similar oscillation was found for the BTZ hole in~\cite{Hodgkinson:2012mr}. 
This oscillation appears to arise from the potential barrier 
in the radial equation~\eqref{eq:4DSchw:radModTort}. 
We show in Appendix \ref{app:barrier} that similar oscillations ensue for 
a static detector in Minkowski spacetime when the field has 
an external potential with a potential wall or a potential barrier. 

Given that the Unruh state represents an outgoing flux of 
radiation from the hole, intuitively one may think that at 
fixed radius $R$ external to the hole, the small-$\omega$ 
up-modes would be unable to escape through the potential 
barrier of~\eqref{eq:4DSchw:radModPhi}. 
Thus, the reader may find it surprising that the transition rate 
of the static detector when the field is in the Unruh state does 
not go to zero as the energy gap goes to zero ---  
implying (by the relation $\tilde{\omega}=E\sqrt{1-2M/R}$ 
that was encountered in Section \ref{sec:4DSchw:static})
that the frequency of the modes is also being taken to zero. 
Because of the theta function 
in~\eqref{eq:4DSchw:staticUnruh_TR_result}, 
the term involving the in-modes is vanishing when $E$ is zero, 
but we show analytically in Appendix \ref{app:A} that the $|\Phup|^2$ 
modes are proportional to 
$\omega^2$ when $\omega\to 0$. 
Hence, this balances the $1/\bigl(E(\expo^{E/T}-1)\bigr)$ 
found in the denominator and leads to a finite transition rate.
\begin{figure}[t]  
\centering
\subfloat[$R=4M$]{\label{fig:static_R4M_HHMinkRind}\includegraphics[width=0.48\textwidth]{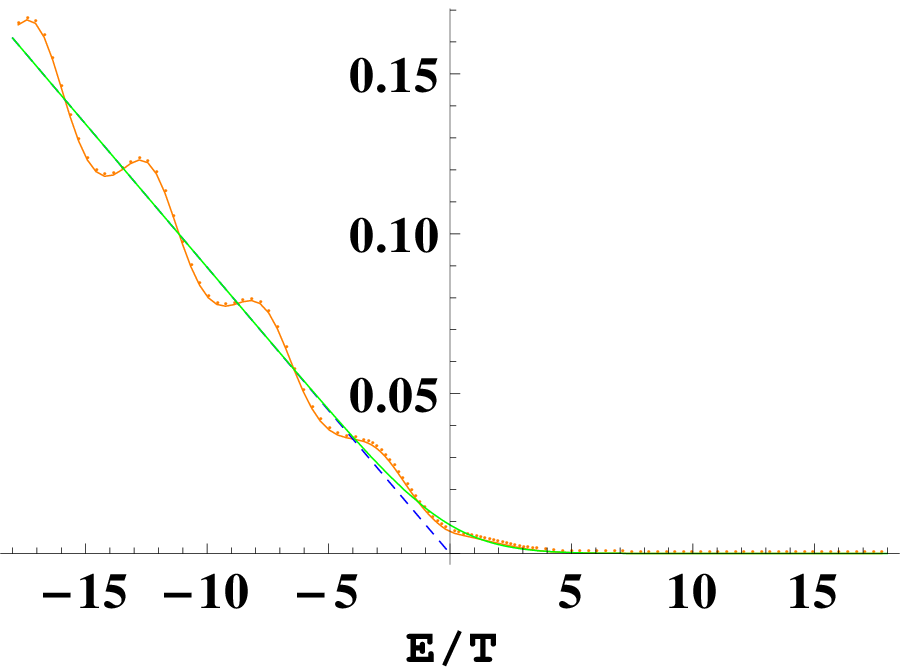}}
\hspace{.5ex}
\subfloat[$R=40M$]{\label{fig:static_R40M_HHMinkRind}\includegraphics[width=0.48\textwidth]{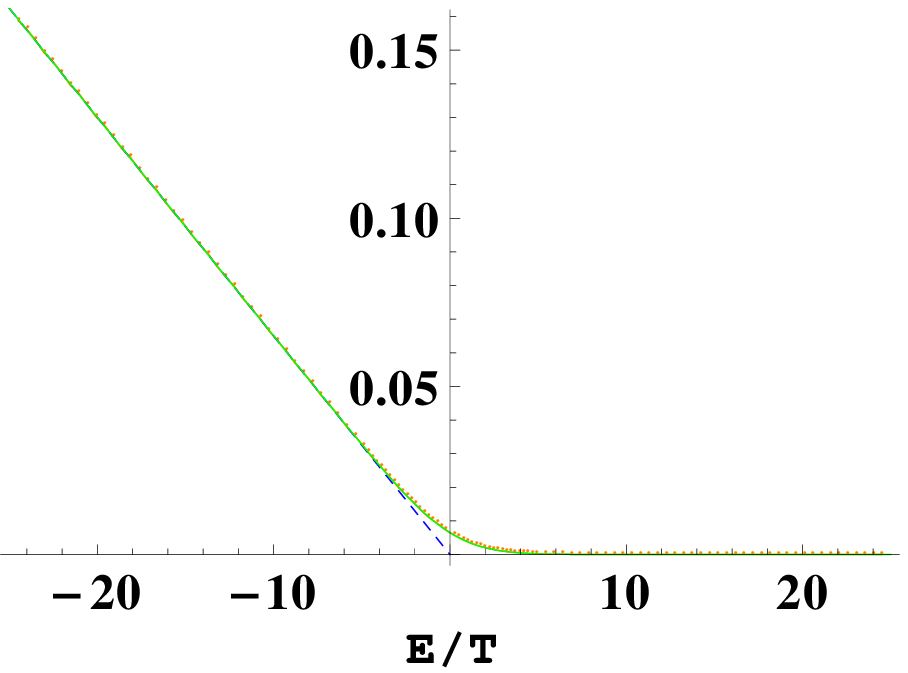}}  
 \caption{$M\dot{\mathcal{F}}$ as a function of $E/T_{\text{loc}}$ for the static detector. Figure showing the results for the Hartle-Hawking state (orange circles), computed from~\eqref{eq:4DSchw:staticHH_TR_result}, alongside the response rate for an inertial detector in $3+1$ Minkowski spacetime (blue dashed), $-\Theta\left(-E\right)E/2\pi$, and the response rate of a Rindler detector (green solid), computed from \eqref{eq:4DSchw:circ:rindRate} with a proper acceleration chosen to be~\eqref{eq:4DSchw:circ:propAccel}.}
\label{fig:static_HHMinkRind}
\end{figure}
\par Figure~\ref{fig:static_HHMinkRind} shows the transition rate of the static detector coupled to a scalar field in the Hartle-Hawking state compared with the transition rate of the inertial detector in 3+1 Minkowski spacetime and a Rindler detector with proper acceleration given by~\eqref{eq:4DSchw:circ:propAccel}. 
First, we see that close to the hole and at large, negative energies the transition rate of the detector coupled to the scalar field in the Hartle-Hawking state, in the black hole spacetime, oscillates about that of the inertial detector, in 3+1 Minkowski spacetime. Second, we observe that as $R$ increases, the Hartle-Hawking rate agrees to an increasing extent with the Rindler detector in flat spacetime. This is to be expected because as one moves further from the black hole the spacetime is asymptotically flat.
\begin{figure}[t]  
\centering
\subfloat[$R=4M$]{\label{fig:static_R4M_ratioHHtoU}\includegraphics[width=0.48\textwidth]{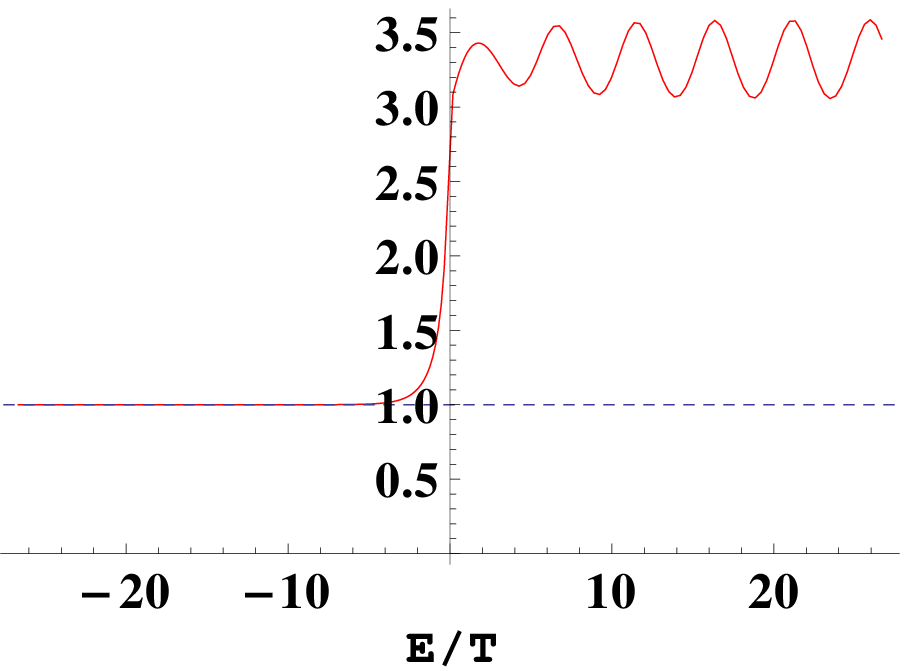}}
\hspace{.5ex}
\subfloat[$R=40M$]{\label{fig:static_R40M_ratioHHtoU}\includegraphics[width=0.48\textwidth]{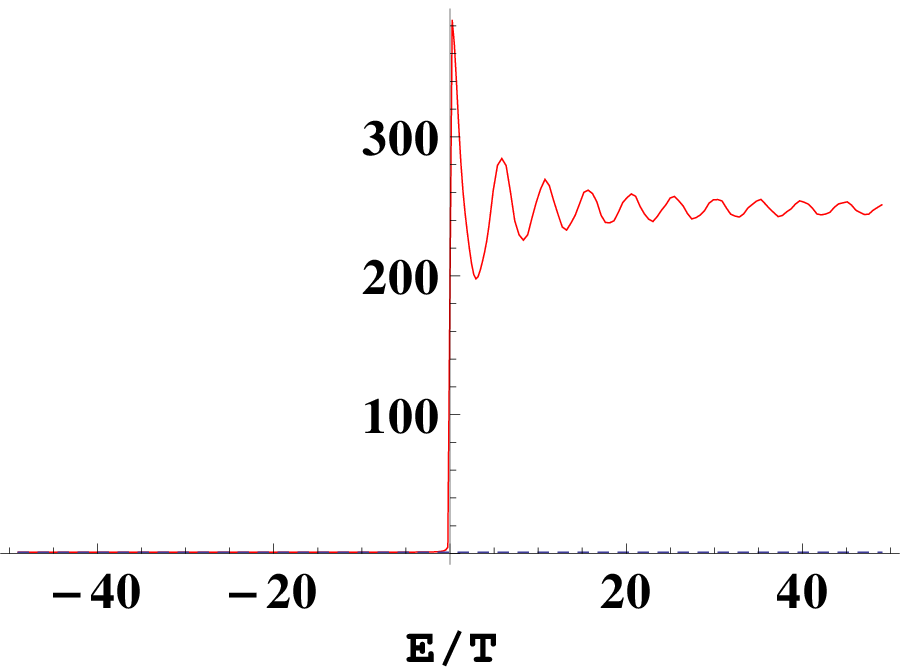}}  
\caption{Ratio of $M\dot{\mathcal{F}}$, as a function of~$E/T_{\text{loc}}$, for the static detector in the Hartle-Hawking state to that of the static detector in the Unruh state. The discontinuity near the origin is caused by the numerical difficulty in computing the modes at small~$\omega$.}
\label{fig:transrate_static_ratioHHtoU}
\end{figure}
\par Figure~\ref{fig:transrate_static_ratioHHtoU} shows the ratio of the transition rate of the static detector coupled to a field in the Hartle-Hawking state to the transition rate of the same detector coupled to a field in the Unruh state. We see that this ratio becomes larger at positive excitation energies and when the radius increases. 
The Unruh state represents a radiating black hole and this radiation will die off by an $r^{-2}$ power law, whereas the Hartle-Hawking state represents a constant heat bath at spatial infinity; therefore, it is to be expected that the ratio between the Hartle-Hawking and Unruh states becomes large as $R\to \infty$. The discontinuity that appears in the 
curves of Figure \ref{fig:transrate_static_ratioHHtoU} is a numerical artefact caused by the fact that solving the ODE \eqref{eq:4DSchw:radModPhi} becomes difficult for small~$\omega$. By the relation $\tilde{\omega}=E\sqrt{1-2M/R}$ that we found in Section~\ref{sec:4DSchw:static}, this means computing the transition rate near $E=0$ is difficult and we did not attempt this.

\begin{figure}[t]  
\centering
\subfloat[$R=4M$]{\label{fig:stat_R4M_KMS}\includegraphics[width=0.48\textwidth]{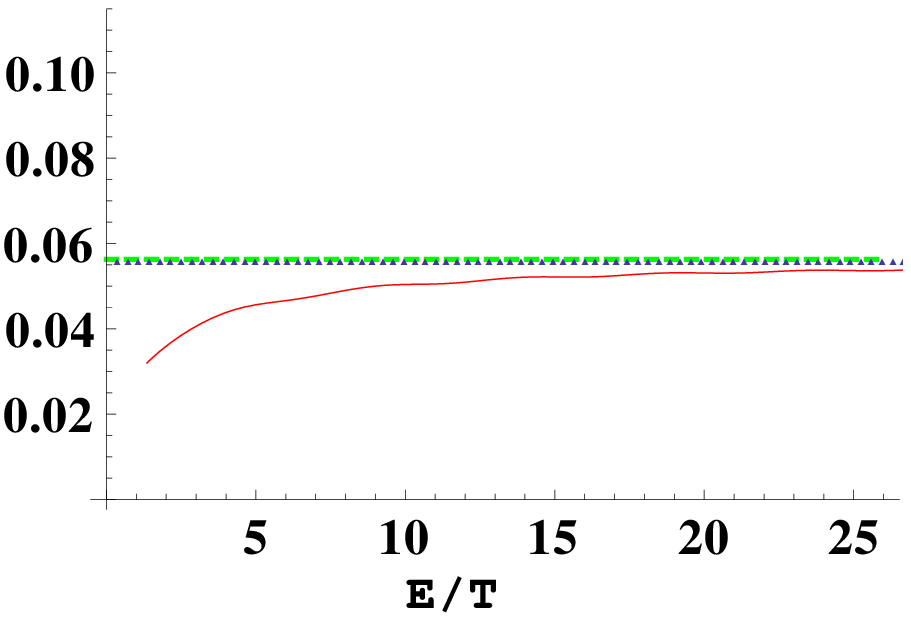}}     
\hspace{.5ex}
\subfloat[$R=40M$]{\label{fig:circ_R40M_KMS}\includegraphics[width=0.48\textwidth]{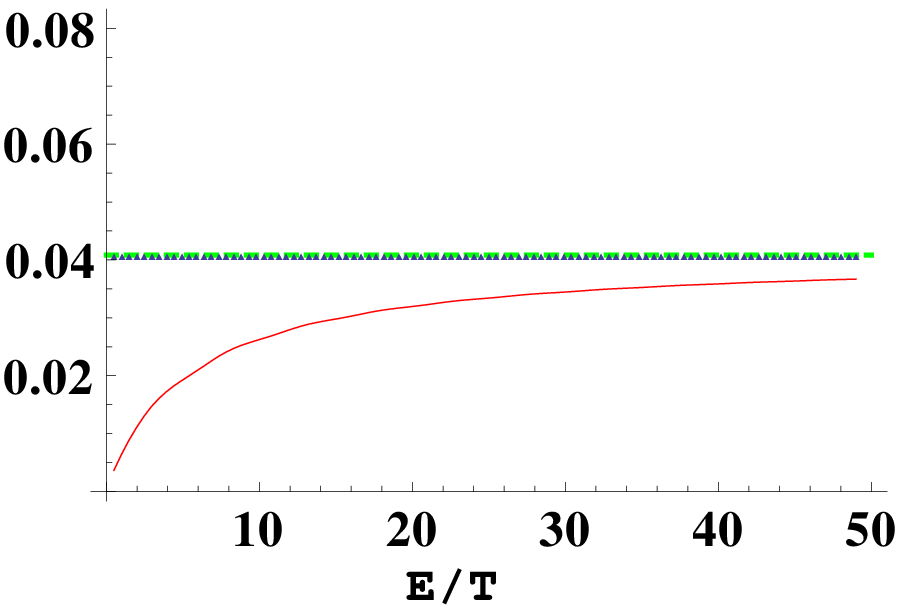}}  
\caption{Figure shows $T_{a}$ \eqref{eq:4DSchw:Ta}
as a function of $E/T_{\text{loc}}$ for a static detector. 
The thick, green, dashed line is the local Hawking temperature~$T_{\text{loc}}$. 
The line with blue triangles is the numerically computed $T_a$ 
for the Hartle-Hawking state, showing agreement with the analytic
result $T_a = T_{\text{loc}}$. 
The solid red curve is the numerically computed $T_a$ for the Unruh state.}
\label{fig:stat_KMS}
\end{figure}

To investigate thermality,  we look at the quantity 
\begin{equation}
T_{a}:=E/\log{\bigl(\mathcal{\dot{F}}(-E)/\mathcal{\dot{F}}(E)\bigr)}\,.
\label{eq:4DSchw:Ta}
\end{equation}
When the KMS condition is satisfied, 
$T_a$ \eqref{eq:4DSchw:Ta} is independent of $E$ and equal to the
temperature. As noted in~\eqref{eq:4DSchw:KMS}, 
this is what happens for the
Hartle-Hawking state, with $T_a = T_{\text{loc}}$. 
Figure~\ref{fig:stat_KMS} shows $T_a$ as a function of
$E/T_{\text{loc}}$ for the Hartle-Hawking state and for the Unruh
state. For the Unruh state, the plot shows that $T_a \to
T_{\text{loc}}$ 
as $E$ increases. This means that in the
limit of a large energy gap, the detector's response in the Unruh state
becomes approximately thermal at the local Hawking temperature~$T_{\text{loc}}$. 

%
%
\subsection[Circular detector]{Circular detector results}
\label{sec:4DSchw:results:circ}
In this section, we present the results obtained for the 
detector on a circular-geodesic in Schwarzschild spacetime. 
These results are computed from the numerical evaluation of the 
transition rates \eqref{eq:4DSchw:circ:HHrateRes}, 
\eqref{eq:4DSchw:circ:BoulwarerateRes} and~\eqref{eq:4DSchw:circ:UnruhrateRes}.

For the circular-geodesic detector's transition rate, we had the double
$\ell$-,~$m$-sum to compute, but as we noted in Section~\ref{sec:4DSchw:circ}, we can
demand that $m\equiv \ell (\text{mod}2)$ to reduce the workload by half. 
We cut off the $\ell$-sum in the transition rate when the contributions at large $\ell$
become negligible. As with the static case, this cut-off is increased as $\omega$ or $R$
increases. Note that for computational efficiency one can take the $\ell$ cut-off of the
$\Phi_{\omega_{-},\ell}$ modes at a significantly lower
value than the $\ell$ cut-off for the $\Phi_{\omega_{+},\ell}$ modes (for both up- and
in-modes).

\begin{figure}[p]  
\centering
\subfloat[$R=4M$]{\label{fig:circ_R4M_3vacs}\includegraphics[width=0.45\textwidth]{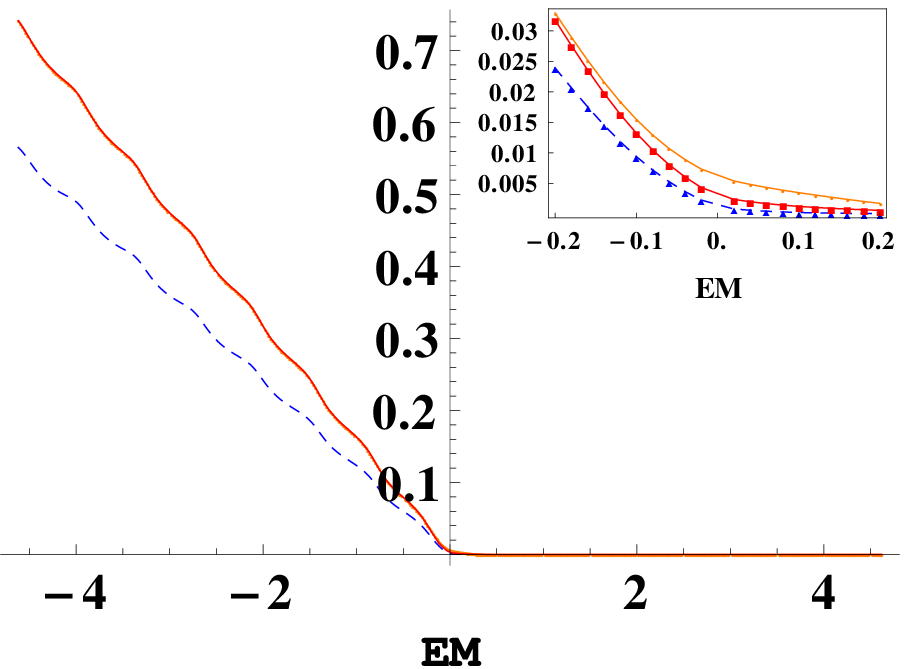}}      
\hspace{1ex}
\subfloat[$R=40M$]{\label{fig:circ_R40M_3vacs}\includegraphics[width=0.45\textwidth]{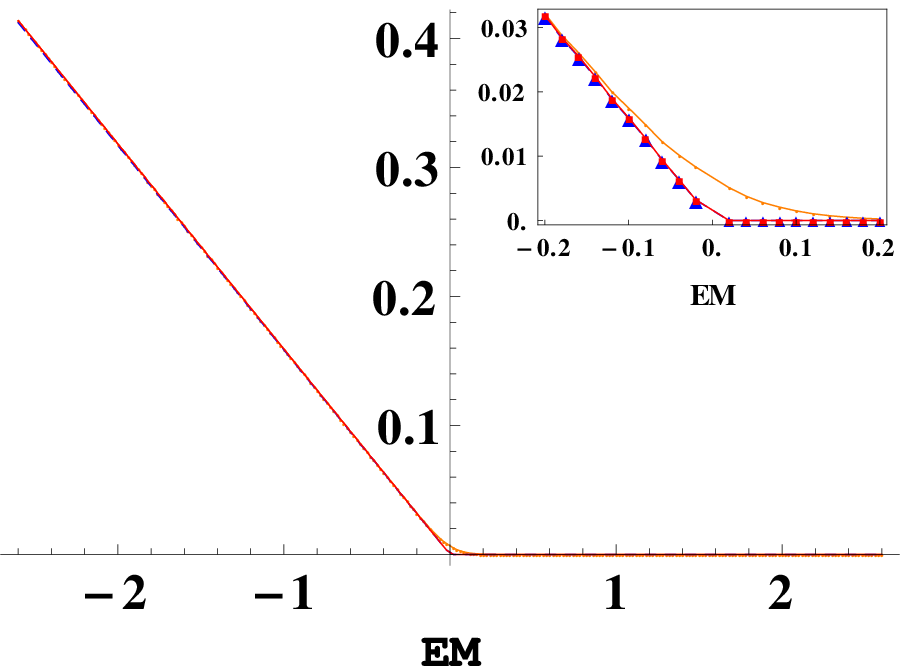}}  
\caption{$M\dot{\mathcal{F}}$ as a function of $EM$ for the circular detector.
The figure shows the transition rate for the Hartle-Hawking state 
(orange circles) computed from~\eqref{eq:4DSchw:circ:HHrateRes}, 
Boulware state (blue dashed) computed from \eqref{eq:4DSchw:circ:BoulwarerateRes} 
and Unruh state (red solid) computed from~\eqref{eq:4DSchw:circ:UnruhrateRes}.}
\label{fig:circ_3vacs}
\end{figure}

\begin{figure}[p]  
\centering
\subfloat[$R=4M$]{\label{fig:circ_R4M_HHRind}\includegraphics[width=0.45\textwidth]{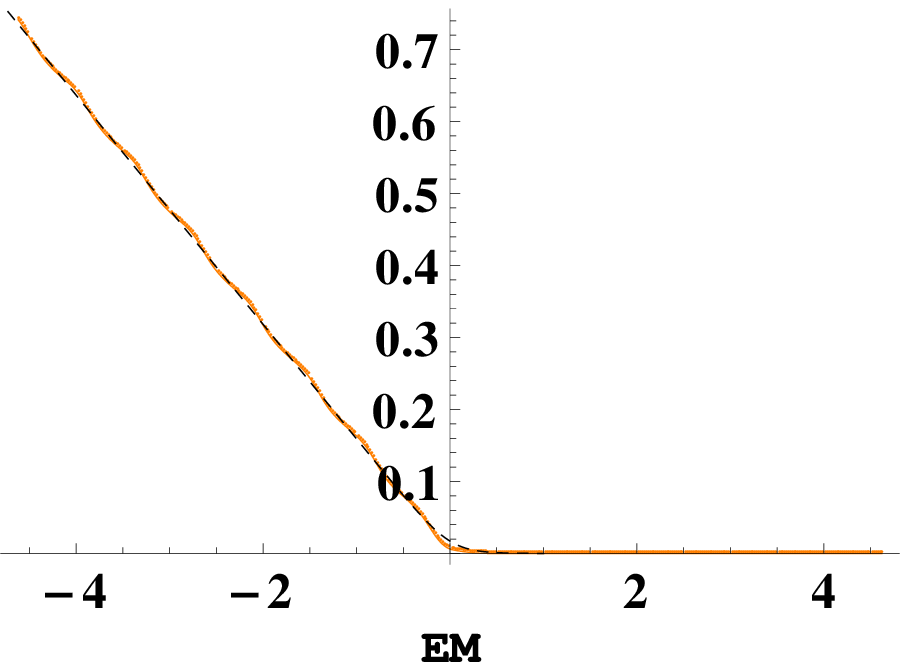}}      
\hspace{1ex}
\subfloat[$R=40M$]{\label{fig:circ_R40M_HHRind}\includegraphics[width=0.45\textwidth]{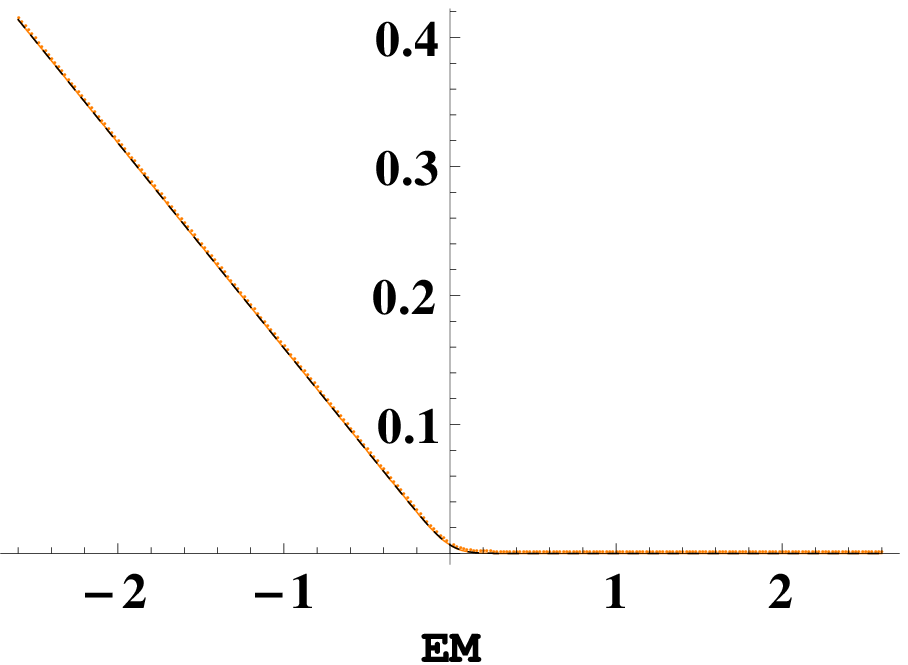}}  
\caption{$M\dot{\mathcal{F}}$ as a function of $EM$ for the circular detector, compared with the Rindler detector with transverse drift. The figure shows the transition rate for the Hartle-Hawking state (orange circles), 
computed from~\eqref{eq:4DSchw:circ:HHrateRes}, alongside the transition rate rate for a Rindler detector with transverse drift (black-dashed). The Rindler rate is computed by substituting the interval~\eqref{eq:4DSchw:circ:interval} into the regulator-free transition rate found in~\cite{louko-satz:profile} and then numerically evaluating.}
\label{fig:circ_HHRind}
\end{figure}

Figure \ref{fig:circ_3vacs} shows the transition rate against the excitation energy of the detector, made dimensionless by the multiplication by the mass of the black hole, $M$. The horizon is at $R=2M$, and we see that as we move away from the horizon, far from the hole at $R=40M$, the transition rates for the Boulware and Unruh states align for negative excitation energies. Below $R=6M$, the circular orbits are unstable but this seems to have no qualitative effect on the transition rate of the detector. Near the hole and for sufficiently large, negative energy gap, the Hartle-Hawking and Unruh rates align, but in the circular case, even at large, negative energies, the Boulware rate does not align with the Hartle-Hawking and Unruh rates.

Figure \ref{fig:circ_HHRind} shows the transition rate of the detector on the Schwarzschild black hole coupled to a scalar field in the Hartle-Hawking state compared with a detector in Rindler spacetime, moving on a Rindler trajectory but drifting with constant velocity in the transverse dimension; that is to say, the trajectory is given by~\eqref{eq:4DSchw:circ:driftTraj}, 
with \eqref{eq:4DSchw:circ:paqRel}, \eqref{eq:4DSchw:circ:q} and~\eqref{eq:4DSchw:circ:p}.  We see that as the radius $R$ increases the agreement becomes better. As $R\to\infty$, the circular detector is becoming asymptotically a static detector, so the agreement should not be surprising considering our results in Section~\ref{sec:4DSchw:results:stat}. Near the hole, at $R=4M$ the transition rate in the Hartle-Hawking state appears to oscillate around that of the drifting Rindler detector when the energy gap is large and negative.

\begin{figure}[p]  
  \centering
  \subfloat[$R=4M$]{\label{fig:circ_R4M_PBC}\includegraphics[width=0.45\textwidth]{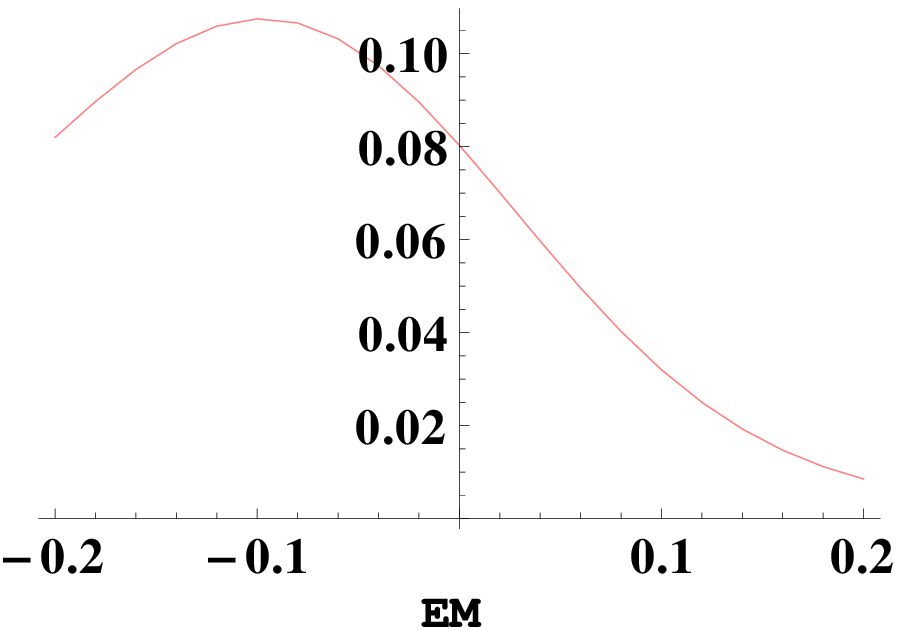}}    
    \subfloat[$R=40M$]{\label{fig:circ_R40M_PBC}\includegraphics[width=0.45\textwidth]{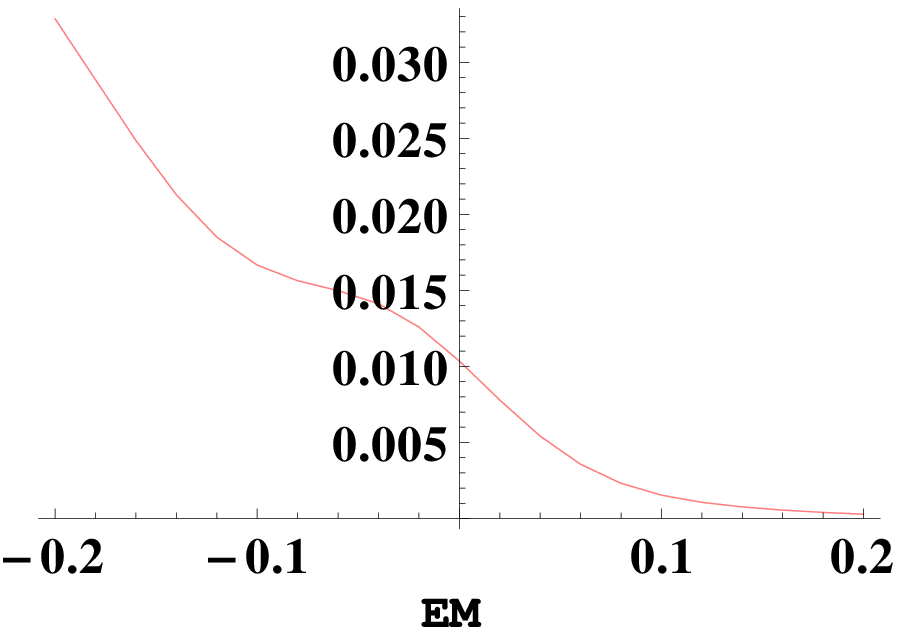}}   
 \caption{Transition rate of a Rindler detector with drift in the transverse direction where the transverse direction has been periodically identified. Computed from~\eqref{eq:4DSchw:circ:n_not0_rate} with $|n|$ cut off at 500.}
\label{fig:transrate_circ_PBC}
\end{figure}

Figure~\ref{fig:transrate_circ_PBC} shows the results that we obtained by making the transverse direction that the drifting Rindler detector's drift occurs in periodic, such that the period matches the period in proper time needed for the circular-geodesic detector, in Schwarzschild spacetime, to complete an orbit. The method of images sum \eqref{eq:4DSchw:circ:n_not0_rate} was cut off at $|n|=500$, by which point the sum had converged. We see by comparing 
Figures \ref{fig:circ_R4M_HHRind} and \ref{fig:circ_R40M_HHRind} with 
Figure \ref{fig:transrate_circ_PBC} that the agreement with the Schwarzschild detector is actually made worse by enforcing periodicity. We note that the oscillation at large, negative energies seen 
in Figure \ref{fig:transrate_circ_PBC} is reminiscent of that seen 
for the co-rotating detector in the BTZ spacetime in~\cite{Hodgkinson:2012mr}.

\begin{figure}[p]  
  \centering
  \subfloat[$R=4M$]{\label{fig:circ_R4M_ratioHHtoU}\includegraphics[width=0.45\textwidth]{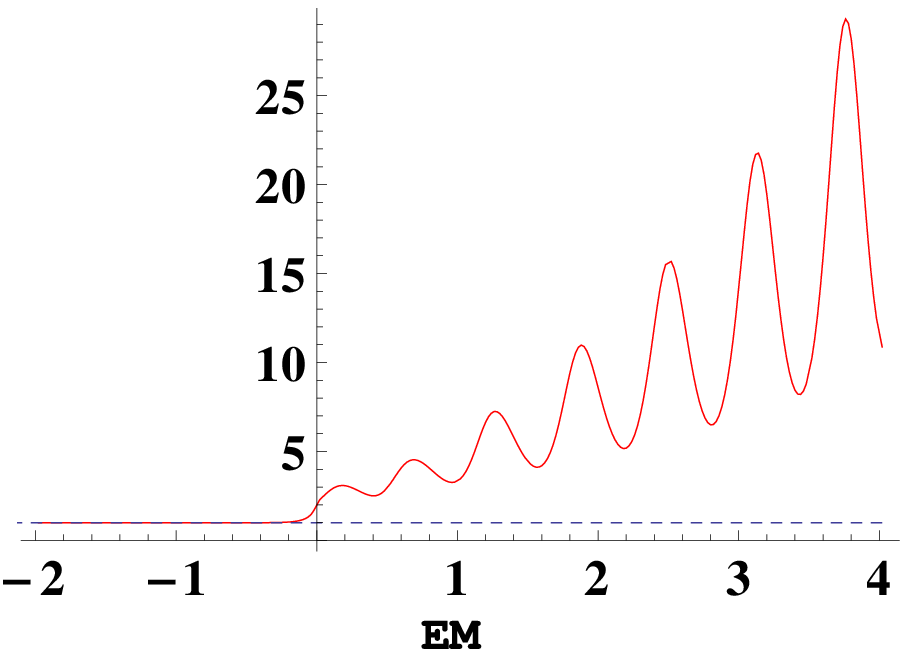}}     
  \subfloat[$R=40M$]{\label{fig:circ_R40M_ratioHHtoU}\includegraphics[width=0.45\textwidth]{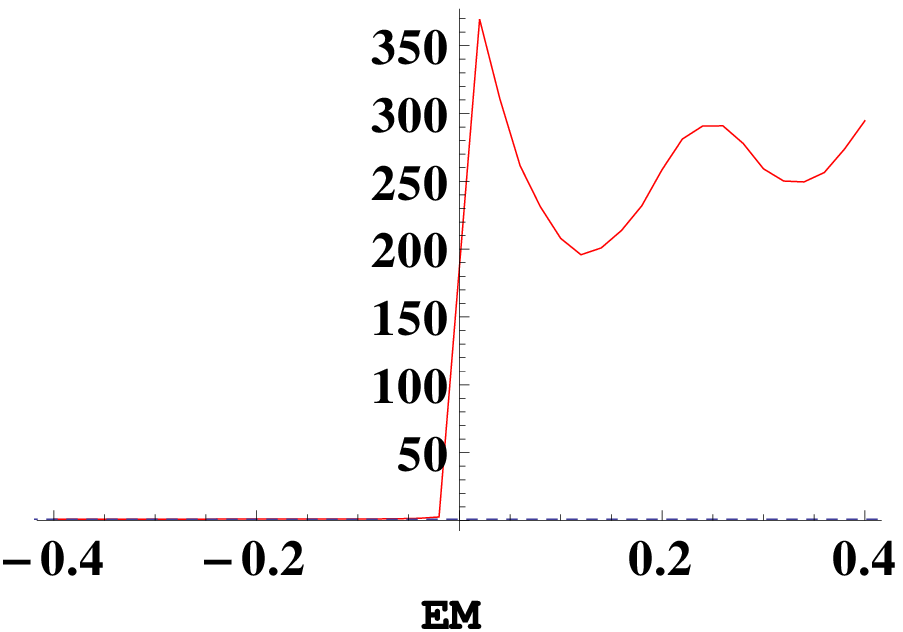}} 
\caption{Ratio of $M\dot{\mathcal{F}}$, as a function of $EM$, for the circular-geodesic detector in the Hartle-Hawking state, to the transition rate of the circular-geodesic detector in the Unruh state. The discontinuity that appears near the origin is a numerical artefact owing to the fact that solving the ODE \eqref{eq:4DSchw:radModPhi} becomes difficult at small~$\omega$.}
\label{fig:transrate_circ_ratioHHtoU}
\end{figure}

Figure~\ref{fig:transrate_circ_ratioHHtoU} shows the ratio of the transition rate of the detector on a circular geodesic coupled to a field in the Hartle-Hawking state, to the transition rate of the circular-geodesic detector coupled to a field in the Unruh state. We see that just like in the static case, this ratio becomes larger at positive excitation energies and when the radius increases.
\begin{figure}[ht!]  
\centering
\subfloat[$R=4M$]{\label{fig:circ_R4M_smerlak}\includegraphics[width=0.48\textwidth]{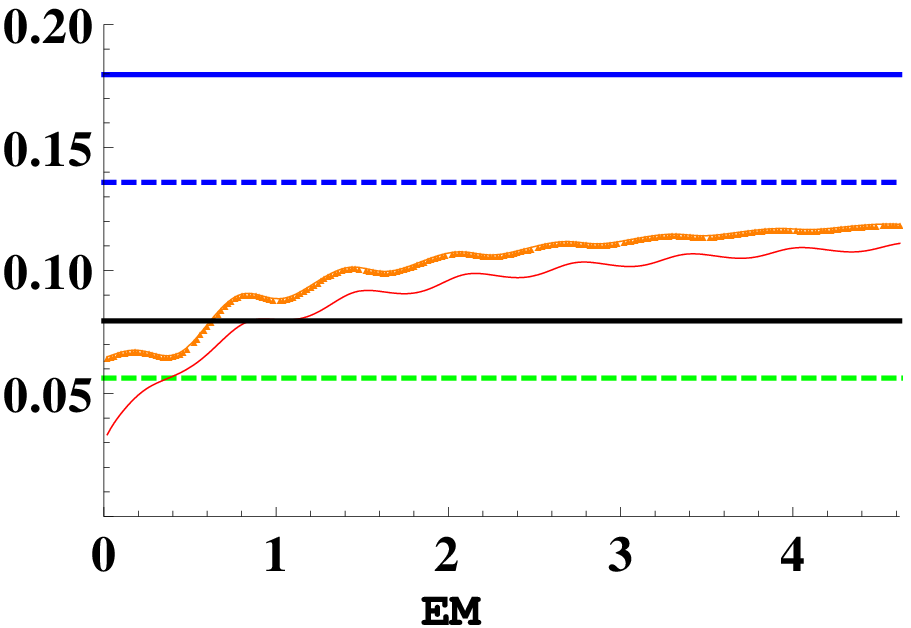}}     
\hspace{.5ex}
\subfloat[$R=40M$]{\label{fig:circ_R40M_smerlak}\includegraphics[width=0.48\textwidth]{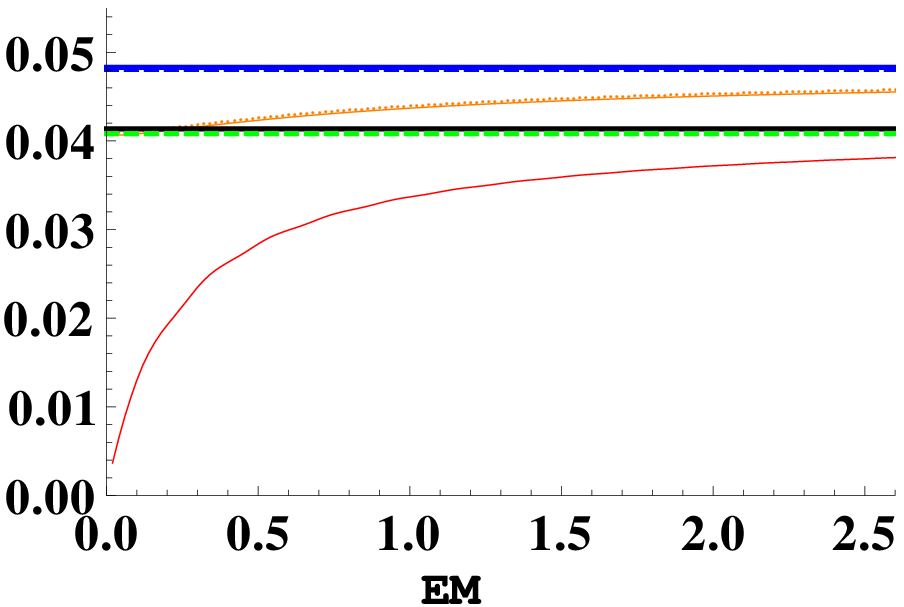}}  
\caption{Figure shows $T_{a}$ \eqref{eq:4DSchw:Ta} as a function of $EM$ for the circular-geodesic detector, in orange circles for the Hartle-Hawking state and in solid red for the Unruh state. 
The horizontal dashed thick green line is the local Hawking 
temperature $T_{\text{loc}}$~\eqref{eq:Tloc-def}, 
and the horizontal solid thick black line is the Doppler-shifted 
local Hawking temperarature $T_{\text{Doppler}}$~\eqref{eq:Tdopp}.
Finally, the horizontal solid and dashed blue lines are obtained 
by shifting $T_{\text{loc}}$ by respectively the factors 
\eqref{eq:ratio-KMSas-Rindler}
and 
\eqref{eq:ratio-KMSas-thermal}
that arise in similar situations involving a drift velocity in 
Minkowski space, as shown in Appendix~\ref{app:stat-Mink-KMS}.}
\label{fig:circ_smerlak}
\end{figure}
\par
Finally, we ask whether the response is thermal in the sense of the KMS property. 
By the discussion at the end of Section~\ref{sec:4DSchw:results:stat}, 
this amounts to examining whether the quantity $T_a$ \eqref{eq:4DSchw:Ta} 
is constant. 

Plots of $T_a$ as a function of $EM$ are shown in 
Figure~\ref{fig:circ_smerlak}. 
Assuming that the range of $EM$ in the plots is representative, 
we see that 
$T_a$ appears to level off as $EM$ increases, both for the 
Hartle-Hawking state and for the Unruh state.
The data does not extend to high enough $EM$ for the 
asymptotic values of $T_a$ to be read off with accuracy,  
but the $R=4M$ plot strongly suggests that
the asymptotic value for each state is higher than the local 
Hawking temperature $T_{\text{loc}}$~\eqref{eq:Tloc-def}, 
and also higher than the Doppler-shifted local 
Hawking temperature, 
\begin{align}
T_{\text{Doppler}} :=
(- \mathsf{U}_{\text{static}} \cdot \mathsf{U}_{\text{circ}})
\, 
T_{\text{loc}}
\ . 
\label{eq:Tdopp}
\end{align}
The $R=40M$ plot supports a similar conclusion for the Hartle-Hawking 
state but remains inconclusive for the Unruh state. 

These results for $T_a$ at a large energy gap 
are similar to what we find in Appendix \ref{app:stat-Mink-KMS} for the 
response of a detector in Minkowski spacetime in three stationary situations 
where the detector has a nonvanishing velocity 
with respect to a family of detectors whose response is exactly~KMS\null. 
The physical explanation for a blueshift above the Doppler 
shift appears to be that while the Doppler shift in \eqref{eq:Tdopp} 
is due to just the time dilation, the 
transition rate at large excitation energies is dominated 
by the most energetic field quanta, and these are seen by the detector 
from a head-on direction and are hence blueshifted 
more than just by time dilation. 
This explanation is consistent with the analysis 
of a circular-geodesic detector in \cite{Smerlak:2013sga}
within a model in which the 
angular dependence of the field is suppressed, where it was found that 
the asymptotic temperature in a state closely resembling the Unruh state is related 
to the local Hawking temperature by just the time dilation Doppler shift factor.

\section{Summary and concluding remarks}
\label{sec:4DSchw:discussion}

In this paper, we have analysed the response of an Unruh-DeWitt
detector coupled to a massless scalar field on the four-dimensional
Schwarzschild black hole using numerical methods.

For the static detector in the exterior region, we analysed the
response when the field was in the Hartle-Hawking, Boulware and Unruh
states. At a variety of radii, the results were presented in the form
of plots of the detector's transition rate, plotted against the
detector's energy gap scaled by the local Hawking temperature.  For
the field in the Hartle-Hawking state, we verified that the response
of the detector was thermal, in the KMS sense, with local temperature
given by $T_{\text{loc}}=1/\bigl(8\pi M \sqrt{1-2M/R}\,\bigr)$, as
known from the complex analytic properties of the Wightman
function. For a static detector and with the field in the Boulware
state, the plots showed that the response of the detector consists
only of de-excitation and that the excitation rate is vanishing; this
is consistent with the fact that the static detector is on an orbit of
the $\partial_t$ Killing vector, where $t$ is the Schwarzschild time
co-ordinate. We also observed from the plots that as the radius
increased, the Boulware and Unruh rates tended to become equal. This
is consistent with the fact that the Unruh rate represents an outgoing
flux of radiation from the hole that diminishes by $r^{-2}$ as the
radius, $r$, tends to infinity, combined with the fact that the
Boulware state tends to the Minkowski vacuum as the radius tends to
infinity. The Hartle-Hawking state represents a thermal heat bath as
the radius tends to infinity, and we plotted the ratio of the
transition rate in the Hartle-Hawking state to the transition rate in
the Unruh state, for the static detector, finding that the ratio of
the excitation rates increases rapidly with radius. We found that for
a large energy gap the transition rate in the Unruh state became
approximately thermal, and the detector recorded the local Hawking
temperature.

We also presented results for a detector on a variety of circular
geodesics, stable and unstable. The results were once again in the
form of plots of the transition rate against the detector's energy
gap, this time scaled to be dimensionless by multiplying by the mass
of the black hole, $M$. Results were presented in the Hartle-Hawking,
Boulware and Unruh states. The stability of the orbit seemed to have
no qualitative effect on the transition rate. The Boulware state in
this case has a non-vanishing excitation component, and this component
increases as the radius decreases. This is consistent with the fact
that at large radius the circular-geodesic detector asymptotes to a
static detector, so the detector becomes approximately on a
$\partial_t$ orbit, but at small radius the detector is no longer on
such an orbit, and there is room for positive energy excitations to
occur. Similarly to the static case, the circular-geodesic plots also
show that as the radius increases, the Boulware and Unruh states tend
to become equal and that the ratio of the Hartle-Hawking rate to Unruh
rate becomes large.

In the limit of a large energy gap, we found evidence 
that the response of a circular-geodesic detector in 
both the Hartle-Hawking state and the Unruh state 
becomes thermal in the KMS sense, in a temperature 
that is higher than the local Hawking temperature, 
by a factor that is genuinely larger than the Dopper 
blueshift factor due to the velocity of the circular 
geodesic with respect to the static detectors. 
This is consistent with the response of a detector in 
three qualitatively similar stationary situations in 
Minkowski spacetime, as we shall show in 
Appendix~\ref{app:stat-Mink-KMS}\null. 
The physical explanation appears to be that the transition rate 
at large excitation energies is dominated by the most energetic field quanta, 
and these are seen by the detector 
from a head-on direction and are hence blueshifted 
more than by the Doppler shift factor that 
accounts for just the time dilation. 
This explanation is consistent with the analysis 
of a circular-geodesic detector in \cite{Smerlak:2013sga}
within a model in which the 
angular dependence of the field is suppressed, where 
the asymptotic temperature in a state closely resembling the Unruh state 
was found to be related 
to the local Hawking temperature by just the time dilation Doppler shift factor. 

Finally for the static detector coupled to a field in the Hartle-Hawking state, a comparison was made to the plot of the transition rate of the Rindler detector in the Minkowski vacuum, with the proper acceleration chosen appropriately. Similarly, for the circular-geodesic detector a comparison was made to a Rindler detector with appropriately chosen proper acceleration, but this time also given a constant velocity drift in the transverse direction; the idea was that this would serve as an analogue to the angular motion of the circular geodesic. The results in both cases showed that as the radius increased, the Hartle-Hawking and Rindler rates aligned. At smaller radius, the Hartle-Hawking rate appears to oscillate about the Rindler (or drifting Rindler) rate.

All the situations analysed were stationary, and we relied on this
stationarity at the outset in order to extract from the formally
divergent total transition probability a finite transition probability
per unit time. While this procedure has a long pedigree~\cite{unruh}, it
would not be applicable in nonstationary situations, such as a detector
falling into a black hole~\cite{Smerlak:2013sga,Barbado:2011dx}. 
For our Schwarzschild Wightman functions that
are given in terms of mode sums, integration over the detection time for
a nonstationary trajectory will no longer collapse the integral
over~$\omega$, and the task of evaluating the transition rate
numerically becomes significantly more involved. In particular, the
Wightman function is divergent at short distances, and while it is known
how the divergent parts come to be subtracted in the expressions for the
transition probability and transition rate~\cite{satz-louko:curved}, the
challenge in numerical work is to implement these subtractions term by
term in a mode sum. For a radially infalling geodesic in Schwarzschild,
a subtraction procedure in the Hartle-Hawking state is 
presented in~\cite{Hodgkinson-thesis},
and a numerical evaluation of the transition
rate is in progress. We hope to report on the results of this evaluation
in a future paper.
\section*{Acknowledgements}
We have benefited from 
discussions with 
numerous colleagues, 
including 
Benito Ju\'arez-Aubry, 
Sanved Kolekar, 
Robb Mann, 
Eduardo Mart\'in-Mart\'inez, 
Eric Poisson, 
Suprit Singh, 
Matteo Smerlak 
and 
Bill Unruh. 
L.H. thanks Bill Unruh for hospitality at the 
University of British Columbia during a 
Universitas 21 Prize Scholarship visit. 
The numerical work was made possible by access to the University of
Nottingham High Performance Computing Facility. 
L.H. was supported by EPSRC through a PhD Studentship
and a PhD Plus Fellowship at the 
University of Nottingham. 
J.L. was supported in part by STFC 
(Theory Consolidated Grant ST/J000388/1). 
A.C.O. acknowledges support from
Science Foundation Ireland under Grant No.\ 10/RFP/PHY2847.
\begin{appendix}
\section{Appendix: Small-$\omega$ behaviour of radial\\ 
up-modes.}
\label{app:A} 
In this appendix, we show that at small $\omega$ 
the up-modes are proportional to~$\omega$. 
The technique overlaps with that in the 
Appendix of \cite{page-masslessrates} 
but using the Whittaker equation \eqref{eq:app:Whittaker} 
allows us to introduce a solution basis that 
remains manifestly regular for 
non-negative integer~$\ell$. 
\par
First, define the dimensionless quantities $x:=r/(2M)-1$ and $k=2M\omega$. 
In terms of these variables, the radial equation~\eqref{eq:4DSchw:radModPhi} reads
\begin{equation}
\left[\frac{d^2}{dx^2}+\frac{(2x+1)}{x(x+1)}\frac{d}{dx}+k^2\left(\frac{x+1}{x}\right)^2-\frac{\ell(\ell+1)}{x(x+1)}\right]\Phi_{\omega\ell}=0\,,
\label{eq:app:radEq}
\end{equation}
where $x>0$. 
%
\par 
In the region $x\gg k+1$, we do a large-$x$ Taylor expansion, 
keeping terms up to and including $O\left(x^{-2}\right)$, resulting in
\begin{equation}
\left[\frac{d^2}{dx^2}+\frac{2}{x}\frac{d}{dx}+k^2\left(\frac{x+1}{x}\right)^2
-\frac{\ell(\ell+1)}{x^2}\right]\Phi_{\omega\ell}=0\,.
\label{eq:app:genSolBigx}
\end{equation}
By writing $\Phi_{\omega\ell}=P/x$ followed by the change of variables $z=2ikx$, 
equation~\eqref{eq:app:genSolBigx} reduces to 
\begin{equation}
\left[\frac{d^2}{dz^2}+\left(-\frac{1}{4}-\frac{ik}{z}+\frac{k^2-\ell(\ell+1)}{z^2}\right)\right]P=0\,.
\label{eq:app:Whittaker}
\end{equation}
This is the Whittaker equation, (13.14.1) of~\cite{dlmf}, with
\begin{equation}
\begin{aligned}
 K&=ik\,,\\
 \mu&=\sqrt{(\ell+1/2)^2-k^2}\,.
 \label{eq:app:WhittakerParams}
\end{aligned}
\end{equation}
\par
Including a suitably chosen phase factor, 
the general solution of~\eqref{eq:app:Whittaker} leads to \cite{dlmf}
\begin{equation}
\begin{aligned}
\Phi_{\omega\ell}&=D_1 \frac{\expo^{-ikx}}{x}(2ikx)^{\mu+1/2}
M\left(\mu+ik + 1/2 ,1+2\mu,2ikx\right)\\
&+D_2\frac{\expo^{-ikx}}{x}(2ikx)^{\mu+1/2}
U\left(\mu+ik + 1/2,1+2\mu,2ikx\right)\,,
\label{eq:app:WhittakerGenSol}
\end{aligned}
\end{equation}
where $D_1$ and $D_2$ are constants. 
\par 
To determine $D_1$ and $D_2$ we compare \eqref{eq:app:WhittakerGenSol} with the $r\to\infty$ asymptotic form of the up-modes, \eqref{eq:4DSchw:asyUp} (now remembering to include the $1/(2M)$ 
normalisation factor of~\eqref{eq:4DSchw:relnAdrianModesNormModes}). After determination of these coefficients, we find that \eqref{eq:app:WhittakerGenSol} reads
\begin{equation}
\begin{aligned}
\Phup&=\frac{\Bup}{(2M)^2}\frac{\Gamma\left(\mu+ik+1/2\right)}{x}\expo^{ik}(2ik)^{-ik}\left(2ikx\right)^{\mu+1/2}\expo^{-ikx}\\
&\times \Big[ \frac{M\left(\mu+ik+1/2,1+2\mu,2ikx\right)}{ \Gamma\left(1+2\mu\right)}\\
&\qquad\qquad-\frac{\expo^{i\pi\left(\mu+ik+1/2\right)}}{\Gamma\left(\mu-ik+1/2\right)}U\left(\mu+ik+1/2,1+2\mu,2ikx\right)\Big]\,.
\label{eq:app:WhittakerGenSolCoeffs}
\end{aligned}
\end{equation}
Next, we look at the limiting form of~\eqref{eq:app:WhittakerGenSolCoeffs} at $1\ll x\ll (\ell+1)/k$; in other words, the limit under consideration is that of large, fixed $x\gg 1$, whilst $k\to 0$.  Hence, expanding in small $kx$, we find, using (13.2.16), (13.14.4) and (13.14.6) of~\cite{dlmf}, that to leading order
\begin{equation}
\begin{aligned}
\Phup&=\frac{\Bup}{(2M)^2x}(-1)^{\ell}\frac{(2\ell)!}{\ell!}k^{-ik}(2ikx)^{-\ell}\\
&=\frac{\Bup}{(2M)^2x}(-1)^{\ell}\frac{(2\ell)!}{\ell!}(2ikx)^{-\ell}\,.
\label{eq:app:LargeXoverlap}
\end{aligned}
\end{equation}
%
%
Finally, using the small-$\omega$ result for the transmission coefficient, $\Bup$, 
given in~\cite{Jensen:1992mv}, we obtain
\begin{equation}
\Phi_{\omega\,,\ell=0}=\frac{1}{2Mx}\left(-2ik\right)\,
\label{eq:app:fin}
\end{equation}
which establishes that for $\ell=0$, $\Phi_{\omega\,,\ell=0} \sim \omega$ as $\omega \to 0$.

\section{Appendix: 
Potential barrier leads to oscillations in the transition rate} 
\label{app:barrier}

In this appendix we discuss two analytically 
solvable systems with a potential barrier. In both systems 
we show that the detector's de-excitation rate depends on the 
detector's energy gap in a way that involves a superposition of 
linear and oscillatory behaviour. 
This is in agreement with the results found numerically   
for the detector in the Schwarzschild spacetime in the main text. 

We consider a scalar field in $(3+1)$-dimensional Minkowski spacetime, 
and we work in a set of standard Minkowski coordinates $(t,x,y,z)$. 
We assume that the field is massless but the wave equation has an external potential 
$V(x)$ that depends on $x$ but not on $t$, $y$ or~$z$, 
such that the spectrum 
of the operator $-(\partial_x^2 + \partial_y^2 + \partial_z^2) + V$ is the positive continuum. 
The wave equation is then separable, and the solutions that have positive frequency 
with respect to $\partial_t$ take the form  
\begin{align}
\phi_{\alpha \kappa_y\kappa_z} (t,x,y,z) 
= \frac{1}{\sqrt{16\pi^3 \omega_\alpha}} \expo^{-i\omega_\alpha t + i\kappa _y y + i \kappa _z z} 
\, u_\alpha(x) 
\label{eq:appbarr-phidef}
\end{align} 
where $\kappa_y$ and $\kappa_z$ are real-valued,
$\alpha$ is a (multi-)index that labels 
the solutions $u_\alpha$ to the one-dimensional wave equation
\begin{align}
- \frac{\text d^2 u_\alpha}{\text{d} x^2} + V(x) u_\alpha(x) = \lambda_\alpha^2 u_\alpha(x)
\label{eq:appbarr-x-eq}
\end{align}
where we may choose $\lambda_\alpha>0$ without loss of generality,  
and $\omega_\alpha$ is the positive solution to the dispersion relation 
\begin{align}
\omega_\alpha^2 =\lambda_\alpha^2+ {\kappa}^2 
\end{align}
with $\kappa=(\kappa _y^2 + \kappa_z^2)^{1/2}$. 
We take the solutions $u_\alpha$ to be normalised as 
\begin{align}
\int u_\alpha(x) u_\beta^*(x) \>\text{d} x= 2\pi \delta_{\alpha\beta}
\ , 
\end{align}  
where $\delta_{\alpha\beta}$ stands for the Dirac delta-function in 
the continuous components of the (multi-)indices $\alpha$ and $\beta$ and 
for a Kronecker delta in any discrete components of the (multi-)indices. 
It follows that the solutions $\phi_{\alpha \kappa_y\kappa_z}$ \eqref{eq:appbarr-phidef}
are then (Dirac) orthonormal in the Klein-Gordon inner product, 
and we may Fock quantise the field in the usual fashion. 

In the vacuum with respect to~$\partial_t$, the transition rate of a 
stationary detector at $x=x_0$ takes the form 
\begin{align}
\dot{\mathcal{F}}(E) &= \int \text{d}\mu(\alpha) 
\int_0^\infty  \text{d} \kappa \kappa 
\, 
\frac{\delta(E+\omega_\alpha) }{4\pi \omega_\alpha} 
\, {|u_\alpha(x_0)|}^2
\nonumber
\\
&= \int \text{d}\mu(\alpha) \int_0^\infty  \text{d} \kappa \kappa 
\frac{\delta\bigl(E+(\lambda_\alpha^2 + {\kappa}^2 )^{1/2}\bigr)}
{4\pi {(\lambda_\alpha^2 + {\kappa}^2 )}^{1/2}} 
\, {|u_\alpha(x_0)|}^2
\nonumber 
\\
&= \frac{1}{4\pi}\int \text{d}\mu(\alpha) \int_{\lambda_\alpha}^\infty  \text{d}s \>\delta(E+s) 
\, {|u_\alpha(x_0)|}^2
\nonumber 
\\
&= \frac{\theta(-E)}{4\pi}  \int \text{d}\mu(\alpha)  \theta( -E -\lambda_\alpha ) 
\, {|u_\alpha(x_0)|}^2
\ , 
\label{eq:appbarr-Fdotgen}
\end{align}
where $\text{d}\mu(\alpha)$ denotes the spectral measure in the (multi-)index~$\alpha$. 

We now apply \eqref{eq:appbarr-Fdotgen} to the well known case of a 
free field in Minkowski space~\cite{takagi}, to the infinite half-space 
potential wall \cite{Langlois:2005nf} and to the 
repulsive P\"oschl-Teller potential. 

\subsection{Free field: no potential barrier}

For the free field in Minkowski space, we have 
$V(x)=0$, $\alpha = k\in\mathbb{R}$, 
$u_k(x) = e^{i k x}$ and $\lambda_k = |k|$. 
From~\eqref{eq:appbarr-Fdotgen}, we obtain 
\begin{align}
\dot{\mathcal{F}}_\text{M} (E) 
&=  \frac{\theta(-E)}{4\pi} 
\int_{-\infty}^\infty \text{d}k \>\theta(-E- |k|)
\nonumber 
\\
&=  -\frac{E}{2\pi} \, \theta(-E)
\ , 
\label{eq:appnobarr-transrate}
\end{align}
which is the well-known result~\cite{takagi}. 

\subsection{Half-space: infinite potential barrier}

For a free field confined to the half-space $x>0$, 
we have $V(x)=0$ for $x\ge0$, and we may think of the 
potential as an infinite wall, so that 
$V(x)=\infty$ for $x<0$. 
With Dirichlet or Neumann boundary conditions at $x=0$, we then have  
$\alpha = k\in\mathbb{R}_+$, 
\begin{align}
u_k(x)= 
\begin{cases}
2 \sin( k x )& \text{for Dirichlet}; 
\\
2 \cos (k x)& \text{for Neumann}, 
\end{cases}
\end{align}
and $\lambda_k = k$. 
For a static detector at $x=x_0>0$, 
\eqref{eq:appbarr-Fdotgen} gives 
\begin{align}
\dot{\mathcal{F}}_\text{wall} (E) 
&=  \frac{\theta(-E)}{4\pi} 
\int_0^\infty \text{d}k \>\theta(-E - k)
\, 
{|u_k(x_0)|}^2
\nonumber 
\\
&= \frac{\theta(-E)}{4\pi} 
\int_0^{-E} \text{d}k
\, 
{|u_k(x_0)|}^2
\nonumber 
\\
&=\frac{1}{2\pi}  \left(-E - \eta \frac{\sin(2 E x_0)}{2x_0} \right)
\theta(-E)
\ , 
\label{eq:appbarr-halfspace}
\end{align}
where $\eta=-1$ for Dirichlet and $\eta=1$ for Neumann. 
This result was obtained in \cite{Langlois:2005nf} by the method of images. 

\begin{figure}[t]  
  \centering
  \subfloat[$\eta=-1$]{\label{fig:app_halfspaceDirichlet}\includegraphics[width=0.49\textwidth]{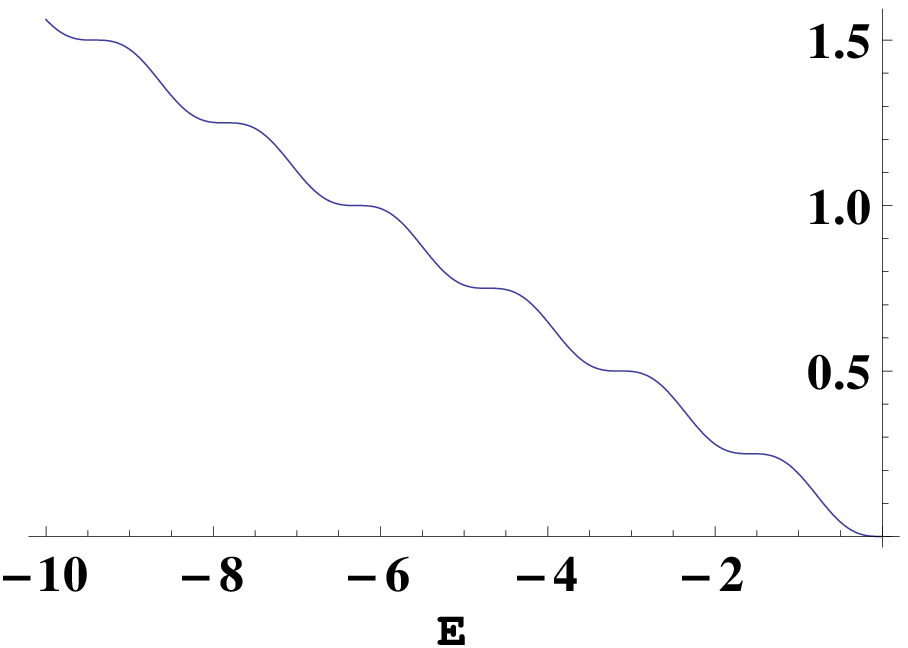}}
  \subfloat[$\eta=1$]{\label{fig:app_halfspaceNeumann}\includegraphics[width=0.49\textwidth]{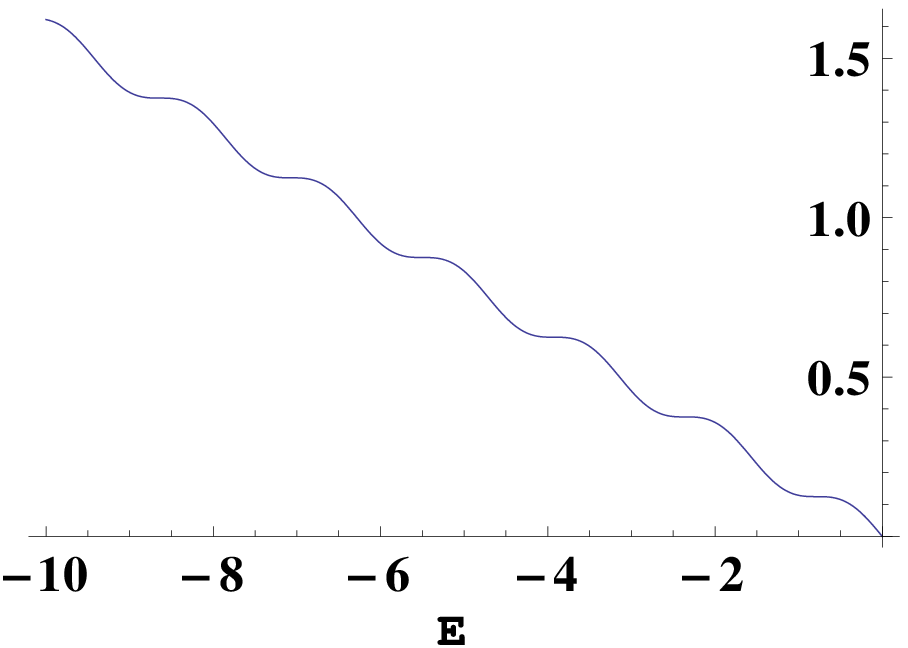}}  
\caption{$\dot{\mathcal{F}}$ as a function of $E$ for a free field in a half-space, 
computed from~\eqref{eq:appbarr-halfspace} at $x_0=2$ for both 
Dirichlet ($\eta=-1$) 
and Neumann ($\eta=1$) 
boundary conditions.}
\label{fig:appFdotWall}
\end{figure}

The transition rate \eqref{eq:appbarr-halfspace} is the superposition of
the Minkowski rate~\eqref{eq:appnobarr-transrate}, linear in~$E$, 
and a term that is oscillatory in $E$ with period~$\pi/x_0$. 
Plots are shown in Figure~\ref{fig:appFdotWall}. 
\subsection{P\"oschl-Teller potential: smooth potential barrier}

As an example of a smooth potential barrier, we consider the repulsive 
P\"oschl-Teller potential, 
\begin{align}
V(x) = \frac{\tfrac{1}{4}+ \mu^2}{\cosh^2 \! x}
\ , 
\label{eq:poschl-teller-potential}
\end{align}
where $\mu^2 > -\frac14$. 
This potential 
is exactly solvable, and it provides a good 
approximation to the potential 
in the Schwarzschild radial equation~\eqref{eq:4DSchw:radModTort}. 

We note in passing that the radial equation for 
massless wave propagation in the 
Nariai spacetime \cite{nariai-first,nariai-second}
can be shown to have exactly the 
P\"oschl-Teller form~\eqref{eq:poschl-teller-potential}. 
This suggests that the Nariai spacetime can 
provide insight into wave propagation in Schwarzschild. 

With the P\"oschl-Teller potential~\eqref{eq:poschl-teller-potential}, 
we may choose the normalised solutions to 
\eqref{eq:appbarr-x-eq} to be 
\begin{subequations}
\label{eq:poschl-teller-solutions}
\begin{align}
u^\text{in}_{k}(x) &= 
\sqrt{\frac{k\pi \sinh k\pi}{\cosh^2(\pi\mu) + \sinh^2(\pi k)}} 
\> \mathrm{P}_{-(1/2) + i \mu}^{i k} ( -\tanh x) 
\ , 
\\
u^\text{up}_{k}(x) &= 
\sqrt{\frac{k\pi \sinh k\pi}{\cosh^2(\pi\mu) + \sinh^2(\pi k)}} 
\> \mathrm{P}_{-(1/2) + i \mu}^{i k} ( \tanh x)
\ , 
\end{align}
\end{subequations}
where $k>0$, $\mathrm{P}$ is 
the associated Legendre function defined with argument on the interval $(-1,1)$ 
by (14.3.1) in~\cite{dlmf}, and $\lambda_k = k$. 
The superscripts ``$\text{in}$'' and ``$\text{up}$'' 
follow the black hole terminology, in the sense that 
$u^\text{in}_{k}$ is proportional to $e^{-ikx}$ at $x\to-\infty$ and 
$u^\text{up}_{k}$ is proportional to $e^{ikx}$ at $x\to\infty$. 
The normalisation in \eqref{eq:poschl-teller-solutions}
can be verified by considering 
the asymptotic behaviour at 
$x\to\pm\infty$~\cite{kleinert-book}.

\begin{figure}[t]  
\centering
\subfloat[$\mu=10$, $x_0=4$]{\label{fig:app_halfspaceDirichlet2}\includegraphics[width=0.49\textwidth]{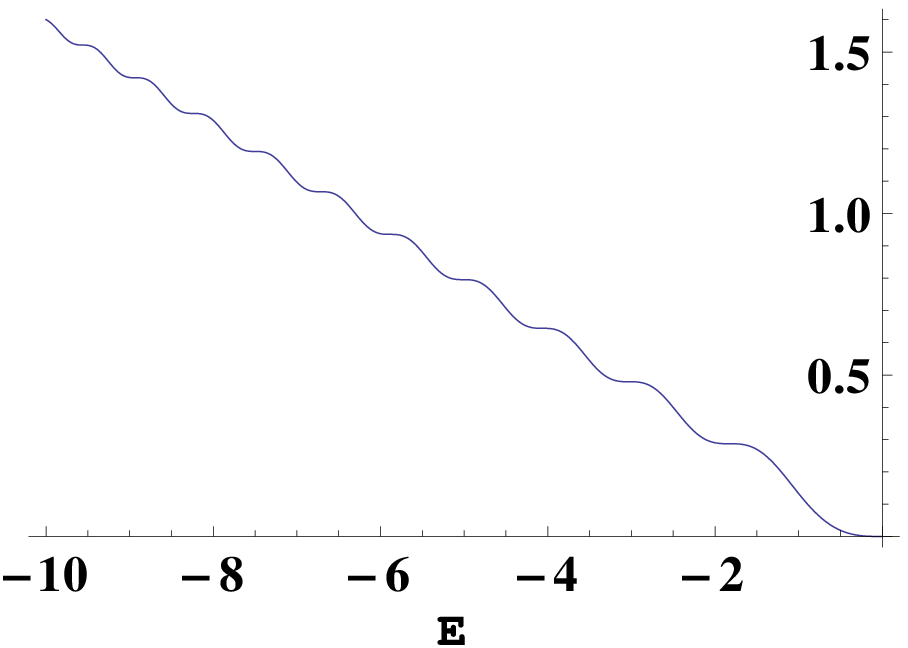}}
\subfloat[$\mu=2$, $x_0=4$]{\label{fig:app_halfspaceNeumann2}\includegraphics[width=0.49\textwidth]{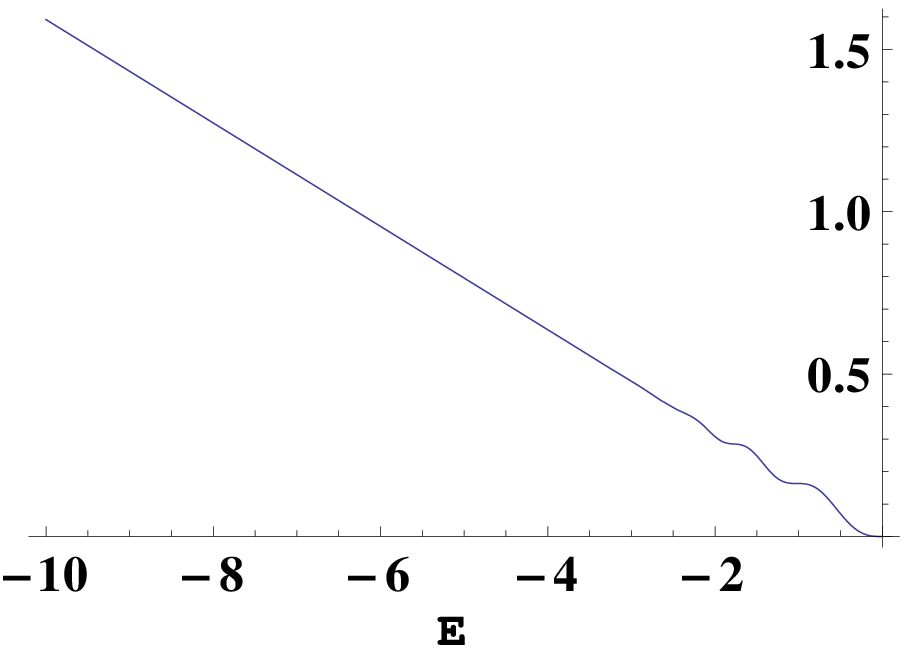}}  
\caption{$\dot{\mathcal{F}}$ as a function of $E$ in the 
P\"oschl-Teller potential, 
computed from \eqref{eq:appbarr-PT}, 
in 
(a) with $\mu=10$ and $x_0=4$,  
and in (b) with $\mu=2$ and $x_0=4$.}
\label{fig:app_FdotPT}
\end{figure}

For a static detector at $x=x_0$, 
\eqref{eq:appbarr-Fdotgen} gives 
\begin{align}
\dot{\mathcal{F}}_\text{PT} (E) 
&=  \frac{\theta(-E)}{4\pi} 
\int_0^\infty \text{d}k \>\theta(-E - k)
\, 
\left( |u^\text{in}_{k}(x_0)|^2
+ 
|u^\text{up}_{k}(x_0)|^2
\right) 
\nonumber 
\\
&= \frac{\theta(-E)}{4\pi} 
\int_0^{-E} \text{d}k
\, 
\left( |u^\text{in}_{k}(x_0)|^2
+ 
|u^\text{up}_{k}(x_0)|^2
\right) 
\ . 
\label{eq:appbarr-PT}
\end{align}
A plot of the transition rate \eqref{eq:appbarr-PT}
is shown in Figure~\ref{fig:app_FdotPT}. 
For small $|E|$ the potential acts as a reflective wall, and the transition rate
superposes oscillatory behavour in $E$ on the linear behaviour of the free field.  
For large $|E|$ the potential wall becomes irrelevant and 
the transition rate asymptotes to that of the free field. 
%
%

%

\section{Appendix: Asymptotic large energy KMS for stationary 
worldlines in Minkowski space}
\label{app:stat-Mink-KMS}

In this appendix we show that the KMS condition holds asymptotically
in the large energy limit for three stationary detector worldlines in
$(3+1)$-dimensional Minkowski spacetime, in situations where the 
detector has a nonvanishing velocity with respect to a 
reference trajectory for which 
the KMS condition holds exactly. 
In all three cases the asymptotic KMS temperature is 
larger than the reference KMS temperature, 
by a factor that exceeds the time dilation Doppler shift 
factor that might be expected on kinematical grounds. 

These analytic results are in qualitative agreement with the 
numerical results found in the main text for the transition rate of a 
detector on a circular geodesic in Schwarzschild. 

We work throughout the appendix in $(3+1)$-dimensional 
Minkowski spacetime. We follow the notation of Section \ref{sec:4DSchw:compRind}
and denote a standard set of Minkowski coordinates by $(t,x,y,z)$. 

\subsection{Rindler with transverse drift in Minkowski vacuum}

We consider a detector on the trajectory~\eqref{eq:4DSchw:circ:driftTraj}, 
\begin{align}
\x(\tau)_{\text{drift}}=\frac{1}{a}
\bigl(
\sinh{\left(q\tau\right)},
\cosh{\left(q\tau\right)},p\tau,0
\bigr)\,,
\label{eq:driftTraj}
\end{align}
where $a>0$, $p>0$, $q = \sqrt{a^2 + p^2}$, 
and $\tau$ is the proper time. 
This trajectory is stationary, 
following an orbit of the Killing vector 
$q(t\partial_x + x \partial_t) + (p/a)\partial_y$. 
Setting $p=0$ yields a Rindler 
trajectory of proper acceleration~$a$: 
compared with this reference Rindler trajectory, 
our trajectory \eqref{eq:driftTraj} 
has a constant drift velocity $v = p/q$. 
A~pair of independent parameters is for example $(a,v)$, 
where $a>0$ and $0<v<1$. 
Alternatively, we may express $v$ in terms of 
the rapidity $\lambda$ by $v = \tanh\lambda$ 
and use the pair $(a,\lambda)$, where $a>0$ and $\lambda>0$. 

We take the field to be in the Minkowski vacuum. 
With switch-on and switch-off pushed to infinity, the stationary
transition rate can be written as~\cite{louko-satz:profile}
\begin{align}
{\dot {\mathcal F}}(E)
= 
{\dot {\mathcal F}}^{\text{inertial}}(E)
+ 
{\dot {\mathcal F}}^{\text{corr}}(E)
\ , 
\label{eq:in-plus-nonin-decomp}
\end{align}
where
\begin{subequations}
\label{eq:in-plus-nonin-formulas}
\begin{align}
{\dot {\mathcal F}}^{\text{inertial}}(E)
&= 
-\frac{E}{2\pi}\Theta(-E)
\ , 
\label{eq:in-base}
\\
{\dot {\mathcal F}}^{\text{corr}}(E)
&= 
\frac{1}{2\pi^2}\int_0^{\infty}\textrm{d}s\cos (E s)
\left( 
\frac{1}{{(\Delta \mathsf{x})}^2} 
+\frac{1}{s^2}
\right) 
\ . 
\label{eq:nonin-correction}
\end{align}
\end{subequations} 
Substituting \eqref{eq:driftTraj} in \eqref{eq:nonin-correction}  
and introducing the new integration variable $z$ by $s = (2/q)z$, we find 
\begin{align}
{\dot {\mathcal F}}^{\text{corr}}(E)
= 
\frac{a^2}{8 \pi^2 q}
\int_{-\infty}^\infty
\textrm{d}z
\, \expo^{2iz|E|/q}
\left(
\frac{1}{(1-v^2) \, z^2}
- 
\frac{1}{\sinh^2 \! z - v^2 z^2}
\right)
\ . 
\label{eq:drift-corr2}
\end{align}

\begin{figure}[t]  
\centering
\subfloat[$v=0.1$]{\label{fig:app_Fcorr_v0p1}\includegraphics[width=0.49\textwidth]{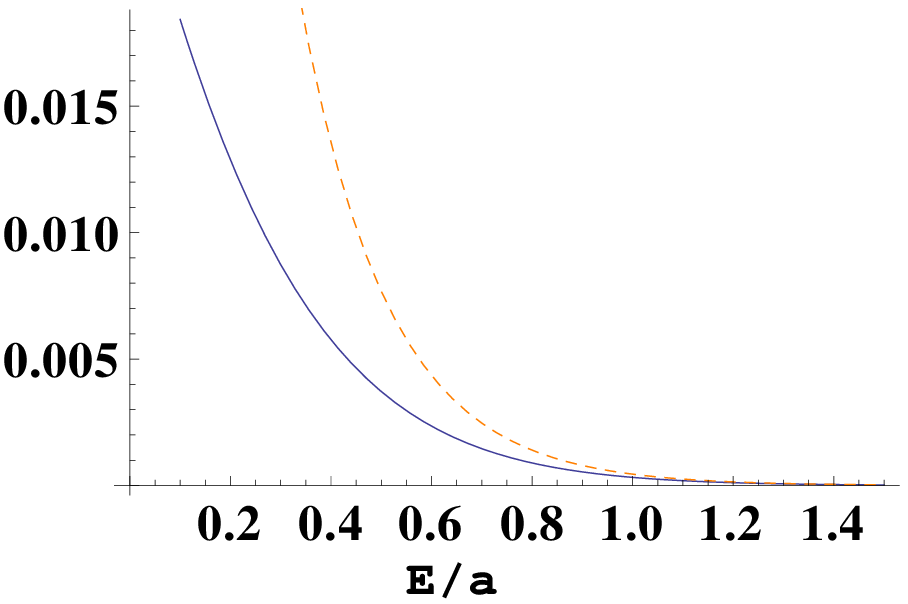}}
\subfloat[$v=0.9$]{\label{fig:app_Fcorr_v0p9}\includegraphics[width=0.49\textwidth]{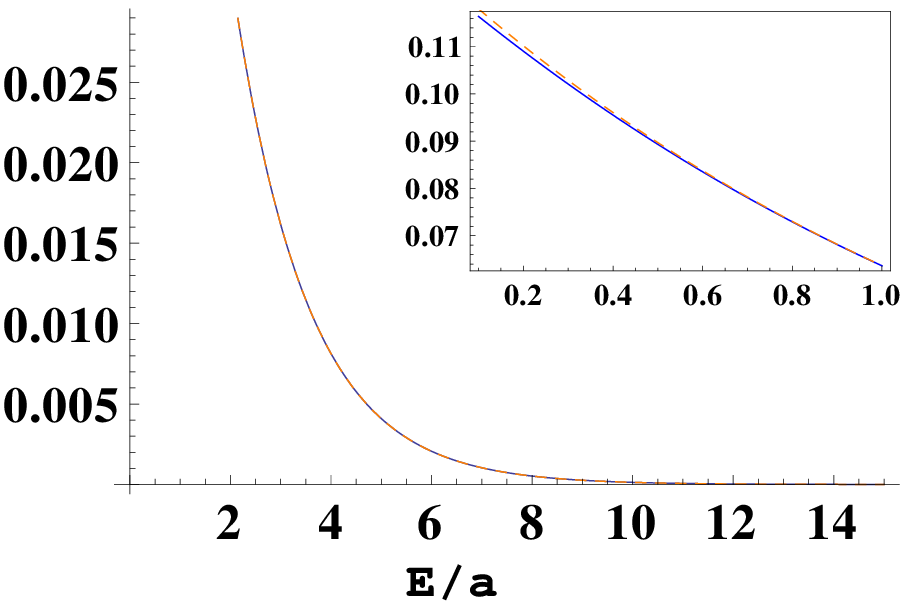}}  
\caption{The solid (blue) lines show the noninertial correction 
${\dot {\mathcal F}}^{\text{corr}}$
\eqref{eq:drift-corr2} to the transition rate of the 
Rindler detector with a transverse drift. 
With 
${\dot {\mathcal F}}^{\text{corr}}/a$ on the vertical axis 
and $E/a$ on the horizontal axis, 
the dimensionful parameter $a$ becomes scaled out and the 
graph depends only on the dimensionless parameter~$v$: 
the two plots show respectively 
(a) $v=0.1$ 
and 
(b) $v=0.9$. 
The dashed (red) curves show the 
asymptotic large energy 
approximation~\eqref{eq:drift-corr-as}.\label{fig:fig:app_Fcorr_coll}}
\end{figure}

To find the leading behaviour of ${\dot {\mathcal F}}^{\text{corr}}(E)$ at
large~$|E|$, we first deform the integration in
\eqref{eq:drift-corr2} to a contour $C$ that passes 
$z=0$ in the upper half of the complex $z$ plane. 
With this contour, the contribution from the first term in
the integrand vanishes and we have 
\begin{align}
{\dot {\mathcal F}}^{\text{corr}}(E)
= 
- \frac{a^2}{8 \pi^2 q}
\int_C
\frac{\expo^{2iz|E|/q} \, \textrm{d}z}{\sinh^2 \! z - v^2 z^2}
\ . 
\label{eq:drift-corr3}
\end{align}
A standard set of contour deformation arguments shows 
that the integral in \eqref{eq:drift-corr3} 
equals $2\pi i$ times the sum of the residues
of the poles in the upper half-plane. 
The dominant contribution at $|E|\to\infty$ 
comes from the pole with the smallest imaginary part, 
which is at $z = iy_+$, where $y_+$ 
is the unique solution to the transcendental equation 
$\sin y= vy$ in the interval $0<y<\pi$. We thus have 
\begin{align}
{\dot {\mathcal F}}^{\text{corr}}(E)
\sim 
\frac{a^2 \exp(- 2|E|y_+/q)}{8\pi q v y_+ (v-\cos y_+)}
\ , \hspace{3ex}
|E|\to\infty
\ . 
\label{eq:drift-corr-as}
\end{align}
A numerical comparison of ${\dot {\mathcal F}}^{\text{corr}}(E)$
and the asymptotic approximation \eqref{eq:drift-corr-as} 
is shown in Figure~\ref{fig:fig:app_Fcorr_coll}. 

From \eqref{eq:in-plus-nonin-decomp},
\eqref{eq:in-base}
and \eqref{eq:drift-corr-as} it
follows that ${\dot {\mathcal F}}(E)$ satisfies at $|E|\to\infty$ the
KMS condition~\eqref{eq:4DSchw:KMS} at the temperature
$T_{\text{R+drift}} = q/(2y_+)$.  
The KMS temperature of the $v\to0$ Rindler 
trajectory is $T_{\text{Rindler}} = 
a/(2\pi)$. 
Hence
\begin{align}
\frac{T_{\text{R+drift}}}{T_{\text{Rindler}}} 
= 
\frac{\pi}{y_+}\cosh\lambda 
\ . 
\label{eq:ratio-KMSas-Rindler}
\end{align}
The ratio \eqref{eq:ratio-KMSas-Rindler} contains 
the time dilation Doppler shift factor~$\cosh\lambda$, 
but also the additional factor~$\pi/y_+$. 
This additional factor is always greater than~$1$, 
tending to $1$ as $v\to0$ and increasing monotonically 
to infinity as $v\to1$.

\subsection{Inertial drift in a thermal bath}

We consider next a detector on the inertial trajectory 
\begin{align}
\x(\tau)= \bigl( 
\tau \cosh\lambda , 
\tau \sinh\lambda , 
0, 0
\bigr)\,,
\label{eq:inertTraj}
\end{align}
where $\lambda>0$ and $\tau$ is the proper time. This trajectory has
the constant drift velocity 
$v=\tanh\lambda$ in the Lorentz-frame defined by the
coordinates.

We now take the field to be in the thermal state at temperature
$T>0$ in the Lorentz-frame defined by the coordinates. 
With switch-on and switch-off pushed to infinity, the
transition rate is stationary, and it is obtained from
\eqref{eq:in-plus-nonin-decomp} and \eqref{eq:in-plus-nonin-formulas}
by first making the replacement 
\begin{align}
\frac{1}{{(\Delta \mathsf{x})}^2} 
\to &
\sum_{n=-\infty}^{\infty}
\frac{1}{ - {(\Delta t + in/T)}^2 + {(\Delta{{\mathbf x})}^2}} 
\nonumber
\\
& = 
\frac{\pi T \sinh(2\pi T |\Delta{{\mathbf x}}|)}{|\Delta{{\mathbf x}}|
\bigl[ 
\cosh(2\pi T |\Delta{{\mathbf x}}|) 
- \cosh(2\pi T \Delta t) 
\bigr]}
\ , 
\label{eq:thermalGreen-gen}
\end{align}
which replaces the Minkowski vacuum by the thermal state, 
and then substituting in the trajectory~\eqref{eq:inertTraj}, 
with the outcome 
\begin{align}
\frac{1}{{(\Delta \mathsf{x})}^2} 
& \to
\frac{\pi T}{2 s \sinh\lambda}
\left[
\coth(\pi \expo^\lambda T s)
- \coth(\pi \expo^{-\lambda} T s)
\right]
\ . 
\label{eq:1overDx2}
\end{align}
Proceeding as in \eqref{eq:drift-corr2} and~\eqref{eq:drift-corr3},
and converting the integral into a sum of the residues in the upper
half-plane, we find
\begin{align}
{\dot {\mathcal F}}(E)
= 
\frac{T}{4\pi \sinh\lambda}
\ln \! \left(
\frac{1 - \exp(- \expo^\lambda E/T)}{1 - \exp(- \expo^{-\lambda} E/T)}
\right)
\ , 
\label{eq:Fdot-drift-bath}
\end{align}
as previously obtained in~\cite{costa-matsas:background}. 
In the limit $\lambda\to0$, \eqref{eq:Fdot-drift-bath} reduces to the
Planckian formula ${(2\pi)}^{-1} E / \bigl(\expo^{E/T}-1\bigr)$. 

At $|E|\to\infty$, ${\dot {\mathcal F}}(E)$ 
\eqref{eq:Fdot-drift-bath} satisfies the KMS
condition \eqref{eq:4DSchw:KMS} at the temperature 
$T_{\text{T+drift}} = \expo^\lambda T$. 
Hence
\begin{align}
\frac{T_{\text{T+drift}}}{T}
= \expo^\lambda
\ . 
\label{eq:ratio-KMSas-thermal}
\end{align}
The ratio \eqref{eq:ratio-KMSas-thermal} 
is equal to the Doppler blueshift factor of the quanta 
that the detector sees head on from the direction of its motion, 
higher than the time dilation Doppler
shift factor $\cosh\lambda$ 
that is experienced by the quanta seen from the 
directions transverse to the motion. 
The transition rate at large excitation energies 
is hence dominated by the most energetic, 
head-on quanta. 
(We thank Eric Poisson for this observation.) 
We note that the ratio \eqref{eq:ratio-KMSas-thermal} 
is not as large as the
ratio~\eqref{eq:ratio-KMSas-Rindler}. 

\subsection{Rotating detector in a thermal bath}

We consider finally a rotating detector. The trajectory is 
\begin{align}
\x(\tau)= \bigl( 
\gamma \tau  , 
R \cos{(\gamma \Omega \tau)} , 
R \sin{(\gamma \Omega \tau)}, 0
\bigr)\,,
\label{eq:circTraj}
\end{align}
where $R>0$, $0<\Omega<1/R$, $\gamma=(1-R^2 \Omega^2)^{-1/2}$, 
and $\tau$ is the proper time. 
The trajectory traces in space a circle of radius~$R$, 
and the angular velocity in the adapted Lorentz frame is~$\Omega$. 
The trajectory is stationary, following an orbit of the Killing vector 
$\partial_t + \Omega(x\partial_y - y\partial_x)$. 
The proper acceleration is $R\gamma^2\Omega^2$. 

We again take the field to be in the thermal state at temperature
$T>0$ in the Lorentz-frame defined by the coordinates. 
With switch-on and switch-off pushed to infinity, the
transition rate is stationary, and it is obtained from
\eqref{eq:in-plus-nonin-decomp} and \eqref{eq:in-plus-nonin-formulas}
by the thermal replacement \eqref{eq:thermalGreen-gen} 
and by substitution of the trajectory \eqref{eq:circTraj}, 
with the outcome 
\begin{align}
\frac{1}{(\Delta\x)^2}
\to
\frac{\pi T \sinh 
\bigl[ 4\pi T R \sin(\gamma\Omega s/2 )
\bigr]}
{2R \sin(\gamma\Omega s/2)
\left\{\cosh
\bigl[
4\pi T R \sin(\gamma\Omega s/2)
\bigr]
-\cosh(2\pi \gamma Ts)\right\}}
\,. 
\label{eq:circ1overDx2}
\end{align}
Proceeding as in \eqref{eq:drift-corr2} and~\eqref{eq:drift-corr3},
we find that the correction to the inertial 
Minkowski vacuum transition rate is given by 
\begin{align}
{\dot {\mathcal F}}^{\text{corr}}(E)
=
\frac{T}{8\pi R \gamma \Omega}\int_C 
\frac{\exp \bigl[i2|E|z/(\gamma \Omega)\bigr]
\sinh(4\pi TR \sin z)
\,\mathrm{d}z}{\sin z \sinh\bigl[2\pi T (R\sin{z}+z/\Omega)\bigr]
\sinh\bigl[2\pi T (R\sin z -z/\Omega )\bigr]}
\,,
\label{eq:rotating-corr}
\end{align}
where we have introduced a new integration variable by $z=\gamma \Omega s/2$ 
and the contour $C$ is along the real axis except for 
passing the pole at $z=0$ in the upper half-plane. 
Note that the integrand in \eqref{eq:rotating-corr}
is nonsingular on $C$ since $R\Omega<1$ and 
the $z\ne0$ zeroes of $\sin z$ 
in the denominator
coincide with simple zeroes in the numerator. 
A~formula equivalent to \eqref{eq:rotating-corr} was obtained
and analysed in certain limits in~\cite{costa-matsas:background}. 

A standard set of contour deformation arguments again shows 
that the integral in \eqref{eq:rotating-corr} 
equals $2\pi i$ times the sum of the residues
of the poles in the upper half-plane. 
The dominant contribution at $|E|\to\infty$ 
comes from the pole with the smallest imaginary part, 
and it can be shown that this pole is on the 
positive imaginary axis in~$z$. 

To identify the dominating pole, let $\rho_+$, $\rho_-$ and
$T_{\text{crit}}$ be the unique positive solutions 
to the transcendental equations 
\begin{subequations}
\label{eq:rot-transc-compilation}
\begin{align}
0 &= 
\sinh \rho_+ - \frac{\rho_+}{R \Omega}
\,,
\label{eq:rhoplus-def}
\\[1ex]
\frac{1}{2RT} &= 
\sinh \rho_- + \frac{\rho_-}{R \Omega}
\,,
\label{eq:rhominus-def}
\\[1ex]
R\Omega 
&= 
4R T_{\text{crit}} \arcsinh\left(\frac{1}{4R T_{\text{crit}}}\right)
\,. 
\end{align}
\end{subequations}
With this notation, the dominating pole is at $z = i \rho_+$ 
when 
$T < T_{\text{crit}}$ and at 
$z = i \rho_-$ 
when 
$T > T_{\text{crit}}$. 
From \eqref{eq:rotating-corr}, we hence have 
\begin{align}
{\dot {\mathcal F}}^{\text{corr}}(E) \sim 
\frac{\exp\bigl[- 2|E|\rho_\pm/(\gamma \Omega) \bigr]}
{8 \pi  R \gamma  \sinh \rho_\pm 
\left(R \Omega \cosh \rho_\pm \mp 1\right)} 
\ , \hspace{3ex}
|E|\to\infty
\ ,
\label{eq:rot-corr-as}
\end{align}
where the upper sign applies for $T < T_{\text{crit}}$ 
and the lower sign applies for $T > T_{\text{crit}}$. 
When $T = T_{\text{crit}}$, 
the two simple poles merge into a dominating 
second-order pole, and the exponential factor 
in \eqref{eq:rot-corr-as} continues to hold 
but the pre-exponential factor gets modified. 

\begin{figure}[t]  
\centering
\includegraphics[width=0.95\textwidth]{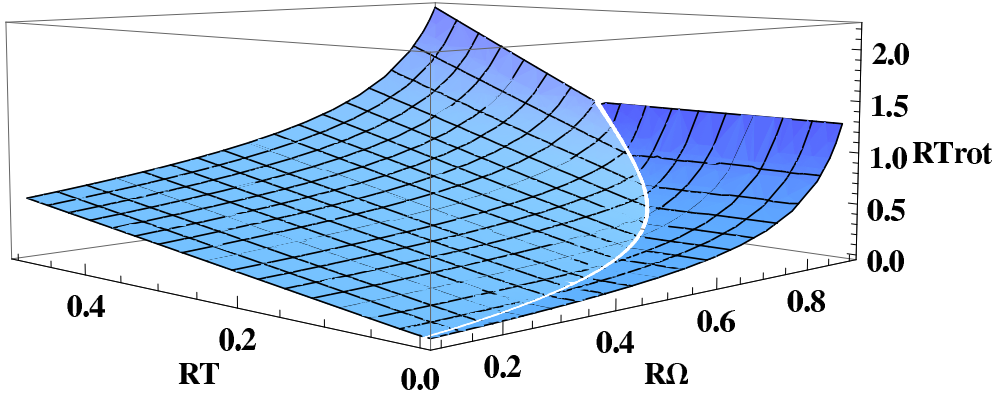}
\caption{The rotating detector's asymptotic temperature 
$T_{\text{rot}}$ \eqref{eq:Trot-endformula} 
is plotted as a function of the ambient temperature $T$ 
and the detector's 
angular velocity~$\Omega$, 
all expressed in units of~$1/R$, 
where $R$ is the radius of the
detector's orbit. 
In the low temperature regime $T_{\text{rot}}$ is independent of~$T$, 
and the transition between the low temperature regime 
and the high temperature regime is 
clearly visible in the plot.  
The limit $\Omega\to0$ at fixed $T$ is in the high temperature regime and 
gives $T_{\text{rot}} \to T$, visible 
in the plot as the straight line at
$\Omega=0$: this is the familiar result for an inertial detector in a
co-moving thermal bath.}
\label{fig:app_3dRot_asyT}
\end{figure}

From \eqref{eq:in-plus-nonin-decomp},
\eqref{eq:in-base} and 
\eqref{eq:rot-corr-as} it
follows that ${\dot {\mathcal F}}(E)$ satisfies at $|E|\to\infty$ the
KMS condition~\eqref{eq:4DSchw:KMS} at the temperature
\begin{align}
T_{\text{rot}} = 
\begin{cases}
{\displaystyle \frac{\gamma\Omega}{2\rho_+}}
& 
\text{for $T < T_{\text{crit}}$}\,,
\\[2ex]
{\displaystyle \frac{\gamma\Omega}{2\rho_-}}
& 
\text{for $T > T_{\text{crit}}$}\,. 
\end{cases}
\label{eq:Trot-endformula}
\end{align}
In the low temperature regime, $T < T_{\text{crit}}$,
$T_{\text{rot}}$ is independent of~$T$: in this regime, 
$T_{\text{rot}}$ is fully determined by the acceleration 
and does not feel the ambient temperature. 
In the high temperature regime,  $T > T_{\text{crit}}$, 
by contrast, 
$T_{\text{rot}}$~depends on both $T$ and~$\Omega$. 
A~plot of $T_{\text{rot}}$ as a function of 
$\Omega$ and $T$
is shown in Figure~\ref{fig:app_3dRot_asyT}. 

We note from \eqref{eq:rot-transc-compilation} and 
\eqref{eq:Trot-endformula} that $R T_{\text{rot}}$ is 
dimensionless and depends only on
the dimensionless combinations $R \Omega$ and~$R T$.
This means that $R$ enters the relations between 
$T_{\text{rot}}$, $\Omega$ and $T$ only as an overall scale. 
The
system can be parametrised by the three independent positive
parameters $(R,\rho_+,\rho_-)$, in terms of which we have 
\begin{subequations}
\label{eq:parametrised}
\begin{align}
R\Omega &=
\frac{\rho_+}{\sinh\rho_+}
\,,
\label{eq:parametrised-ROmega}
\\
RT &=
\frac{1}{2 \left( \sinh\rho_- + \rho_- \rho_+^{-1} \sinh\rho_+ \right)}
\,,
\\
RT_{\text{crit}} &= 
\frac{1}{4\sinh\rho_+}
\,,
\\
RT_{\text{rot}} &= 
\begin{cases}
{\displaystyle \frac{1}{2\sqrt{\sinh^2\rho_+ - \rho_+^2}}}
& 
\text{for $\rho_- > \rho_+$ \ ($T < T_{\text{crit}}$)}\,,
\\[4ex]
{\displaystyle \frac{\rho_+}{2\rho_-\sqrt{\sinh^2\rho_+ - \rho_+^2}}}
& 
\text{for $\rho_- < \rho_+$ \ ($T > T_{\text{crit}}$)}\,,
\end{cases}
\label{eq:Trot-endformula-par}
\end{align}
\end{subequations}
where the low temperature regime $T < T_{\text{crit}}$ 
occurs for $\rho_->\rho_+$ and the
high temperature regime $T > T_{\text{crit}}$ occurs for $\rho_-<\rho_+$. 

We wish to compare $T_{\text{rot}}$ to the ambient
temperature~$T$. From \eqref{eq:Trot-endformula} 
and~\eqref{eq:parametrised}, we have 
\begin{align}
\frac{T_{\text{rot}}}{T} 
= 
\begin{cases}
{\displaystyle  \gamma 
\left(
\frac{\rho_-}{\rho_+} + \frac{\sinh\rho_-}{\sinh\rho_+}
\right)}
& 
\text{for $\rho_- > \rho_+$ \ ($T < T_{\text{crit}}$)}\,,
\\[3ex]
{\displaystyle \gamma 
\left(
1 + \frac{\rho_+ \sinh\rho_-}{\rho_-\sinh\rho_+}
\right)}
& 
\text{for $\rho_- < \rho_+$ \ ($T > T_{\text{crit}}$)}\,. 
\end{cases}
\label{eq:ratio-rot-KMSas-in}
\end{align}
The ratio \eqref{eq:ratio-rot-KMSas-in} contains 
the expected time dilation Doppler
shift factor~$\gamma$, but also an additional factor that is always
greater than unity, taking 
values in 
the interval $(1,2)$ in the high
temperature regime and in the half-line $(2,\infty)$ in the low
temperature regime. Note that this additional factor depends 
not just on the detector's trajectory but also on~$T$, 
even in the high-temperature regime $T > T_{\text{crit}}$. 

As a final observation, we consider the limit $T\to0$, in which the
field is in the Minkowski vacuum, and we compare $T_{\text{rot}}$ 
to the Unruh temperature of a 
Rindler trajectory with the same value of proper acceleration, 
$T_{\text{Rindler}} = R \gamma^2\Omega^2/(2\pi)$. 
From~\eqref{eq:parametrised}, we obtain 
\begin{align}
\left.
\frac{T_{\text{rot}}}{T_{\text{Rindler}}} 
\right|_{T=0}
= 
\frac{\pi \sqrt{\sinh^2 \! \rho_+ - \rho_+^2}}{\rho_+^2} 
\,,
\hspace{3ex}
\label{eq:ratio-rot-KMSas-Rind}
\end{align}
where $\rho_+$ is determined by $R\Omega$
from~\eqref{eq:parametrised-ROmega}. 
The ratio \eqref{eq:ratio-rot-KMSas-Rind}
takes values on the
half-line $(\pi/\sqrt{3}, \infty)$, 
asymptoting to $\pi/\sqrt{3}$ 
in the ultrarelativistic limit $R\Omega\to1$ 
and to $\infty$ in the inertial limit 
$R\Omega\to0$. This disagreement between $T_{\text{rot}}$ and
$T_{\text{Rindler}}$ highlights the qualitative differences between
linear acceleration and circular acceleration in Minkowski
vacuum~\cite{Letaw:1980yv,Grove:1983rp,Muller:1995vk}. 

\end{appendix}

\end{document}